\documentclass[aps,prd,10pt,nofootinbib,OliveGeqsecnum,showpacs,showkeys,superscriptaddress,preprintnumbers,altaffilletter,floatfix,twocolumn]{revtex4-2}

\usepackage{graphicx,caption,subcaption,bm,color,hyperref}
\usepackage{amsmath,amssymb}
\usepackage{mathtools} 
\usepackage{dsfont} 
\usepackage{multirow} 
\usepackage{float} 
\usepackage{mathrsfs} 
\usepackage{tensor} 
\usepackage{booktabs}
\usepackage{caption}
\usepackage{ragged2e}

\hypersetup{colorlinks, linkcolor={blue}, citecolor={red}, urlcolor={teal}}

\definecolor{amaranth}{rgb}{0.9, 0.17, 0.31}
\definecolor{palatinateblue}{rgb}{0.15, 0.23, 0.89}
\definecolor{brightpink}{rgb}{1.0, 0.0, 0.5}
\definecolor{brightgreen}{rgb}{0.14, 0.84, 0.72}
\definecolor{mediumgreen}{rgb}{0.22, 0.67, 0.59}
\definecolor{darkgreen}{rgb}{0.25, 0.5, 0.46}
\definecolor{verydarkgreen}{rgb}{0.22, 0.33, 0.31}

\hypersetup{ linktoc=all,
    colorlinks, linkcolor={mediumgreen},
    citecolor={mediumgreen}, urlcolor={darkgreen}
}

\graphicspath{{images/}}

\newcommand{\be}{\begin{equation}}
\newcommand{\ee}{\end{equation}}
\newcommand{\ba}{\begin{eqnarray}}
\newcommand{\ea}{\end{eqnarray}}

\begin{document}

\title{Cosmographic analysis of sign-switching dark energy}

\author{Mariam Bouhmadi-López}
\email{mariam.bouhmadi@ehu.eus}
\affiliation{IKERBASQUE, Basque Foundation for Science, 48011, Bilbao, Spain}
\affiliation{Department of Physics  \& EHU Quantum Center, University of the Basque Country UPV/EHU, P.O. Box 644, 48080 Bilbao, Spain}

\author{Beñat Ibarra-Uriondo}
\email{benat.ibarra@ehu.eus}
\affiliation{Department of Physics \& EHU Quantum Center, University of the Basque Country UPV/EHU, P.O. Box 644, 48080 Bilbao, Spain}

\begin{abstract} 

Inspired by the well-studied $\Lambda_{\rm s}$CDM model, we propose and investigate a class of dynamical dark energy models that evolve from a negative cosmological constant, transitioning to a positive value at low redshifts. Specifically, we introduce a dark energy framework based on a generalised ladder-step function for the cosmological constant. Furthermore, we present two additional models in which the cosmological constant undergoes a smooth sign change at a specified redshift. We provide a detailed discussion of the construction and theoretical properties of these models, analysing their background cosmological evolution using a cosmographic approach and the statefinder hierarchy parameters. Our results are compared with those obtained for the standard $\Lambda$CDM and $\Lambda_{\rm s}$CDM models. We also perform a careful analysis of the types of singularities that may arise in these models due to a sign change in the cosmological constant. Notably, we show that a continuous sign change removes the sudden singularity present in the $\Lambda_{\rm s}$CDM model, replacing it with a milder $w$-singularity.

\end{abstract}

\maketitle

\section{Introduction}\label{intro}

The late-time acceleration of the expansion of the universe, first observed in 1998 \cite{SupernovaSearchTeam:1998fmf}, has since been robustly confirmed \cite{SupernovaCosmologyProject:1998vns,Bahcall:1999xn}. Although the underlying cause of this acceleration remains unknown, the simplest approach to account for it is the introduction of a cosmological constant, giving rise to the $\Lambda$CDM model. While this model successfully reproduces the large-scale behaviour of the Universe, several problems and tensions have emerged, highlighting unresolved issues within this framework, which we briefly summarise below.

The cosmological constant as a mean to describe the late-time acceleration of the Universe introduces a new energy scale, $\rho_v \approx 10^{-47}~\text{GeV}^4$, which is remarkably small compared to typical energy scales in particle physics~\cite{RevModPhys.61.1,Sahni:1999gb}. Moreover, the so-called coincidence problem arises~\cite{Carroll:2000fy,Padmanabhan:2002ji}, as there is no natural explanation for why the present-day energy densities of matter and dark energy (DE) are of the same order, $\Omega_{\mathrm{m},0} \sim \Omega_{\mathrm{d},0}$, without invoking fine-tuning of the initial conditions. In addition to the coincidence problem, recent observations have revealed a tension in the determination of the Hubble constant, $H_0$. Measurements at early times tend to yield higher values than those inferred from late-time observations~\cite{DiValentino:2021izs,Poulin:2018,Kamionkowski:2022pkx}. A similar discrepancy is found in the measurement of $S_8$~\cite{DiValentino:2021izs,Perivolaropoulos:2021jda,Abdalla:2022yfr,Bamba:2012cp}. 
Furthermore, there exists the so-called neutrino tension, i.e. cosmological neutrino mass bounds conflict with laboratory measurements ~\cite{Lesgourgues:2012uu,daFonseca:2023ury}.
Due to these tensions and open questions, numerous alternatives to the standard $\Lambda$CDM cosmological model have been explored over the past two decades. These include models based on scalar fields, such as quintessence~\cite{PhysRevD.37.3406} and k-essence~\cite{PhysRevD.63.103510},  axion-like DE models~\cite{Kamionkowski:2014zda,Emami:2016mrt,Chiang:2025qxg}, perfect fluids~\cite{Kamenshchik:2001cp}. We can also find various modified theories of gravity~\cite{CANTATA:2021asi}, such as bigravity theories~\cite{Kobayashi:2019hrl},   kinetic gravity braiding (KGB) \cite{Deffayet:2010qz,Pujolas:2011he,BorislavovVasilev:2022gpp,BorislavovVasilev:2024loq}, $f(\mathcal{R})$ gravity \cite{Sotiriou:2008rp,Capozziello:2011et,Nojiri:2010wj,Nojiri:2017ncd}, $f(\mathcal{T})$ \cite{Bengochea:2008gz,Ferraro:2006jd,Cai:2015emx} and  $f(\mathcal{Q})$ theories~\cite{BeltranJimenez:2018vdo,BeltranJimenez:2019tme,Ayuso:2020dcu,Boiza:2025xpn,Ayuso:2025vkc}  among others.

Motivated by the previously studied $\Lambda_{\rm s}$CDM model \cite{Akarsu:2022typ}, we investigate cosmological scenarios that involve a transition from a negative to a positive cosmological constant in the late-time Universe ($z < 2$ \cite{Akarsu:2019hmw}). This $\Lambda_{\rm s}$CDM model offers a potential resolution to the $H_0$ tension \cite{Akarsu:2023mfb}, thereby addressing certain shortcomings of the $\Lambda$CDM scenario. Moreover, the presence of a negative cosmological constant during cosmic expansion is particularly appealing, as it evokes the AdS/CFT correspondence \cite{Maldacena:1997re}, and arises naturally in string theory and string-theory-inspired supergravity frameworks.
It is also worth noting that a positive cosmological constant, such as that in the $\Lambda$CDM model, presents significant theoretical challenges, beyond those already mentioned. For instance, formulating quantum field theory  on a de Sitter background ($\Lambda > 0$), as well as obtaining vacuum solutions with $\Lambda > 0$ within string theory, remains a notoriously difficult task \cite{Obied:2018sgi,Maldacena:2000mw,Kachru:2003aw}. These issues could potentially be circumvented if a negative cosmological constant existed in the Universe.

There already exists a substantial body of literature examining scenarios with a negative cosmological constant. In the context of an inflationary Universe, numerous studies explore inflationary models involving multiple AdS vacua \cite{Piao:2004me,Li:2019ipk,Vazquez:2018qdg}. Within the framework of Early Dark Energy (EDE), some models propose the presence of AdS vacua around the epoch of recombination as a means of alleviating the $H_0$ tension \cite{Planck:2018vyg}. Regarding post-recombination modifications to the $\Lambda$CDM model, several studies \cite{Sahni:2014ooa,Mortsell:2018mfj,Poulin:2018zxs,LinaresCedeno:2021aqk,Souza:2024qwd,Gomez-Valent:2024ejh} suggest that cosmological data are fully consistent with—or may even favour—a negative cosmological constant at high redshifts. Such scenarios also arise in modified gravity theories, including Brans–Dicke theory \cite{Faraoni:1998qx} and quadratic bimetric gravity \cite{Mortsell:2018mfj}.

The $\Lambda_{\rm s}$CDM model, originally proposed in \cite{Akarsu:2021fol} and based on the findings of the graduated dark energy (gDE) model \cite{Akarsu:2019hmw,Acquaviva:2021jov}, has been shown to alleviate the $H_0$ and $S_8$ tensions by increasing the value of the Hubble parameter to better fit observational data \cite{Akarsu:2021fol,CosmoVerse:2025txj,Souza:2024qwd,Escamilla:2025imi}. From a mathematical standpoint, at redshifts $z < z_\dagger$, where $z_\dagger$ denotes the sign-switching redshift, $\Lambda_{\rm s}$CDM is indistinguishable from $\Lambda$CDM, exhibiting a de Sitter (dS)-like cosmological constant (CC) asymptotically. However, for $z > z_\dagger$, a minimal modification is introduced, resulting in an AdS-like CC. 
Although this modification does not affect the early Universe, as the DE density is negligible at that epoch, the change around the sign-switching redshift, which is close to the matter–dark energy equality, is sufficient to address most of the current cosmological tensions. Nevertheless, from a phenomenological perspective, the impact of this modification is only significant at $z \lesssim z_\dagger$, with observational constraints estimating the free parameter to be $z_\dagger < 2$ \cite{Akarsu:2019hmw,Akarsu:2021fol,Akarsu:2022typ,Planck:2018vyg,Akarsu:2023mfb}.  Current  studies  tend to favour values around $z_\dagger \sim 1.7$ \cite{Escamilla:2025imi}.
Recent developments have explored extensions of the $\Lambda_{\rm s}$CDM framework. A particularly compelling example is the $\Lambda_{\rm s}$CDM$^+$ model, which emerges within a string-theoretic context \cite{Anchordoqui:2023woo,Anchordoqui:2024gfa,Anchordoqui:2024dqc,Soriano:2025gxd}. In this extension, the transition from negative to positive DE density is not imposed by hand but arises dynamically via Casimir forces acting in the bulk; i.e. the higher dimensional space-time. This demonstrates how revisiting classical scenarios within broader theoretical frameworks can uncover previously overlooked mechanisms capable of realising such transitions in a natural and consistent manner.
Furthermore, taking inspiration from models akin to the $\Lambda_{\rm s}$VCDM model investigated in \cite{Akarsu:2024qsi,Akarsu:2024eoo}, which features a smooth transition in contrast to the abrupt sign-switch examined in \cite{Planck:2018vyg,Akarsu:2019hmw}, we introduce three variations of the mentioned scenario: one with a ladder-like cosmological constant, another motivated by a smooth step function, and a third that follows the profile of an error function. These are compared against the abrupt sign-switching model.
The primary objective of these models is to explore phenomenological alternatives that exhibit smoother behaviour or transitions, rather than to provide a fundamental resolution to the cosmological constant problem. In future work, we aim to embed this behaviour within a fundamental theory that may help to alleviate current cosmological tensions. Although the differences between the models considered here are primarily reflected in their cosmographic parameters, several observational avenues exist to test them. Cosmological fits can indicate whether data prefer any specific model, while both model-dependent \cite{Luongo:2024fww} and model-independent \cite{Velazquez:2024aya} techniques allow constraints on the cosmographic quantities computed in this work. Additional signatures could arise from perturbative analyses using $f\sigma_8$ measurements \cite{pert} or from the study of singularities predicted by these scenarios \cite{Paraskevas:2024ytz}, which may impact bound cosmic structures.

This paper is organised as follows. In Sec. \ref{sec2}, we briefly present and review DE models from a phenomenological perspective. In particular, we review the $\Lambda_{\rm s}$CDM model and introduce three new models, comparing all of them in Subsec. \ref{sec2e}. In Sec. \ref{sec3}, we analyse each of these models through a cosmographic/statefinder analysis. In Sec. \ref{sec5}, we conclude by highlighting key aspects of our work. In Appendix \ref{sing},
{we conduct a careful analysis of the types of singularities that may arise in these kinds of models due to a change in the sign of the cosmological constant. We demonstrate that a continuous change in the sign of the cosmological constant eliminates the sudden singularity (type II) present in the $\Lambda_{\rm s}$CDM model, replacing it with a milder $w$-singularity (type V) \cite{Paraskevas:2024ytz}.}
Finally, in Appendices \ref{appendixb}, \ref{appendixc} and \ref{appendixd}, we present some useful equations for Sec. \ref{sec3}.

\section{Beyond the \texorpdfstring{$\Lambda_{\rm s}$}{Lambda{s}}CDM Model \label{sec2}}

In this section, we briefly review the abrupt sign-switching DE model, denoted as $\Lambda_{\rm s}$CDM, which we shall refer to as model (A). Subsequently, we introduce three new models inspired by it, namely: (B) Ladder-like DE model (L$\Lambda$CDM), (C) Smooth Step DE model (SSCDM), and (D) Error-function DE model (ECDM). For each model, we first present the equation of state (EoS) for DE that forms the basis of its formulation, ensuring that the background evolution remains closely aligned with that of the $\Lambda$CDM model at the lowest redshifts and up to the present epoch. 


Let us begin by considering a homogeneous and isotropic universe described by the Friedmann-Lema\^itre-Robertson-Walker (FLRW) metric. For a spatially flat universe ($k=0$), the Friedmann and Raychaudhuri equations take the form:
\begin{equation}
    H^2=\frac{\kappa^2}{3} \rho, \quad \frac{\ddot{a}}{a}=-\frac{\kappa^2}{6}(\rho+3p).
\end{equation}

Here, $H \equiv \frac{\dot{a}}{a}$ denotes the Hubble parameter, where a dot represents differentiation with respect to cosmic time, $\{\dot{\ } \} \equiv \frac{d}{dt}$. The parameter  $\kappa^2$ is defined as  $8\pi G$, where $G$ denotes the gravitational constant. The total energy density and pressure of the Universe are denoted by $\rho$ and $p$, respectively. In this work, we consider a multi-fluid Universe composed of radiation, matter (comprising cold dark matter and baryons), and DE. Accordingly, the total energy density and pressure can be decomposed as
\begin{equation}
    \rho=\rho_{\mathrm{r}}+\rho_m+\rho_{\mathrm{d}} \quad \text { and } \quad p=p_{\mathrm{r}}+p_{m}+p_{\mathrm{d}},
    \label{densityenergy}
\end{equation}
where the subscripts $r$, $m$, and $d$ denote radiation, matter, and DE, respectively.  We shall also neglect interactions between the individual matter components. As a result, each fluid $A = r, m, d$ satisfies the usual conservation equation:
\begin{equation}
    \dot{\rho}_A+3 H\left(\rho_A+p_A\right)=0.
\end{equation}

The individual EoS parameter for each fluid reads 
\begin{equation}
    w_{\mathrm{r}}=\frac{p_{\mathrm{r}}}{\rho_{\mathrm{r}}}=\frac{1}{3}, \quad w_{\mathrm{m}}=\frac{p_{\mathrm{m}}}{\rho_{\mathrm{m}}}=0, \quad w_{\mathrm{d}}=\frac{\rho_{\mathrm{d}}}{p_{\mathrm{d}}}.
\end{equation}

It can be shown that the DE density evolves with  redshift as:
\begin{equation}
    \rho_{\mathrm{d}}(z)=\rho_{\mathrm{d},0}  \text{    exp}\left(3\int_0^z\frac{1+w_\mathrm{d}(z')}{1+z'}dz'\right),
\end{equation}
while the matter and radiation components evolve as $\rho_\mathrm{m}=\rho_\mathrm{m,0} (1+z)^3$ and $\rho_\mathrm{r}=\rho_\mathrm{r,0} (1+z)^4$.
This enables us to write the Friedman equation taking all fluids into consideration, as shown below:
\begin{equation}
\begin{aligned}
    \frac{H(z)^2}{H_0^2}= & \Omega_{\mathrm{r},0}(1+z)^4+\Omega_{\mathrm{m},0}(1+z)^3 \\
    & +\Omega_{\mathrm{d},0}\frac{\rho_{\mathrm{d}}(z)}{\rho_{\mathrm{d},0}},
    \end{aligned}
    \label{eq6}
\end{equation}
where, from now on, a $0$-subscript denotes the present value of a given quantity. We can as well  write Eq.~(\ref{eq6}) in terms of the fractional energy densities of each component; i.e.,
\begin{equation}
\Omega_{\mathrm{r}}=\frac{\rho_{\mathrm{r}}}{\rho_c}, \quad \Omega_{\mathrm{m}}=\frac{\rho_{\mathrm{m}}}{\rho_c}, \quad \Omega_{\mathrm{d}}=\frac{\rho_{\mathrm{d}}}{\rho_c},
\label{Omega}
\end{equation}
where $\rho_c=\frac{3H^2}{\kappa^2}$. 

The DE EoS parameter will be introduced later for each individual model. From Eqs. (\ref{densityenergy}) and (\ref{Omega}) we can obtain the total EoS parameter, $w$, from the individual ones:
\begin{equation}
w \equiv \frac{p}{\rho}=\Omega_\mathrm{r} w_\mathrm{r}+\Omega_\mathrm{m} w_\mathrm{m}+\Omega_\mathrm{d} w_\mathrm{d}.
\label{eq:totalEoS}
\end{equation}

In the following subsections, we first revisit the model previously studied in Ref.~\cite{Akarsu:2022typ}, followed by the introduction of three new models, which will be the focus of this work. For the rest of this section, we will only assume a matter and DE filled universe, disregarding the presence of radiation, as it is subdominant at lower redshifts.

\subsection{Abrupt sign-switching \texorpdfstring{$\Lambda_{\rm s}$}{Lambda{s}}CDM model \label{sec2-A}}

We begin by reviewing a one-parameter extension of the $\Lambda$CDM model, known as the $\Lambda_{\rm s}$CDM model \cite{Akarsu:2021fol}. This model is characterised by a DE component given by a negative cosmological constant that switches sign at late times, i.e.,
\begin{equation}
\Lambda_\mathrm{s}(z) = \Lambda \text{ sgn}(z_{\dagger,s} - z),
\end{equation}
where $z_{\dagger,s}$ denotes the redshift at which the sign-switching occurs, and $\Lambda$ represents the 
value of the cosmological constant at present where the model reduces to the well known
 $\Lambda$CDM scenario. The function $\text{sgn}(z_{\dagger,s} - z)$ denotes the sign function, which returns $1$ for positive arguments and $-1$ for negative ones.

Since the DE density remains constant over time, the EoS parameter remains the same as in $\Lambda$CDM, i.e., \mbox{$w_d = -1$,} for most of the time, except at the sign-switching point. In fact, by invoking the conservation of the energy density of $\Lambda_\mathrm{s}(z)$, it can be demonstrated that:
\begin{equation}
    w_{\textrm{d,s}}=-1-\frac{\delta(z_{\dagger,s}-z)}{3 \textrm{ sgn}(z_{\dagger,s}-z)}(1+z),
\end{equation}
where $\delta(z_{\dagger,s}-z)$ stands for the Dirac delta function. Therefore, we observe that the EoS parameter is ill-defined at the time of sign-switching, leading to a singularity that will be reanalysed later. Assuming a matter and DE filled universe, the evolution of the scale factor is given by \cite{Akarsu:2021fol}
\begin{equation}
  \begin{aligned}
&a(t)= \begin{cases}A^{\frac{1}{3}} \sin ^{\frac{2}{3}}\left[\frac{3}{2} \sqrt{\frac{\Lambda}{3}} t\right], & \text { for } t \leq t_{\dagger,s}, \\ A^{\frac{1}{3}} \sinh ^{\frac{2}{3}}\left[\frac{3}{2}\sqrt{\frac{\Lambda}{3}} t+B\right], & \text { for } t \geq t_{\dagger,s},\end{cases}\\
\end{aligned}
\end{equation}

where
\begin{equation}
\begin{aligned}
& A=\sinh ^{-2}\left[\frac{3}{2}\sqrt{\frac{\Lambda}{3}} t_0+B\right], \\
& B=\operatorname{arcsinh}\left[\sin \left(\frac{3}{2} \sqrt{\frac{\Lambda}{3}} t_{\dagger,s}\right)\right]-\frac{3}{2} \sqrt{\frac{\Lambda}{3}} t_{\dagger,s}.
\end{aligned}  
\label{eq:scale-factorle}
\end{equation}

In such a model, although the scale factor is continuous, the Hubble parameter is not, as shown in Fig. (\ref{fig:hubble}), and is smaller than in the $\Lambda$CDM model.  To derive this solution, we have normalised the scale factor such that $a(t_0)=1$. Under this condition, general relativity implies, through the Friedmann equations, that this solution satisfies $A=\Omega_{\mathrm{m},0}/\Omega_{\mathrm{d},0}=\kappa^2\rho_{\mathrm{m},0}/\Lambda$ (c.f. Eq (\ref{eq:scale-factorle})). For further details on this model, please refer to Ref. \cite{Akarsu:2021fol}.

\subsection{Ladder-like DE model (L\texorpdfstring{$\Lambda$}{Lambda}CDM) \label{sec2-B}}
The first type of model we present is a natural generalisation of the $\Lambda_{\rm s}$CDM model, in which the effective cosmological constant increases gradually in discrete steps. Here, each step represents an increase in the cosmological constant via a ``small'' discontinuous function, where the degree of smallness depends on the number of assumed steps.
For simplicity, we shall assume that all the steps are equal. This type of model can be divided into two cases, depending on the number of steps assumed, as even number of steps introduce a step at which the DE density becomes null.

\subsubsection{Even ladder-like DE model (Le\texorpdfstring{$\Lambda$}{Lambda}CDM) \label{sec2b} }

This is a two-parameters extension of the $\Lambda$CDM model, in which  we consider an initial and final redshifts, denoted respectively by $z_{i,le}$ and $z_{f,le}$. Those parameters  determine when the cosmological constant starts growing from a minimum negative value at $z_{i,le}$ to a maximum  positive value at $z_{f,le}$. The growth of the cosmological constant is implemented via an $N$-step ladder function, with a controllable step length. In this model, the number of steps $N=20$ is chosen to be even, and the DE density is given by:
\begin{equation}
  \begin{aligned}
&  \Lambda_\mathrm{le}(z)=\Lambda \left[1-\frac{1}{10}\sum_{n=1}^{20} \mathcal{H}(z_n-z)\right],
\end{aligned}  
\label{densityladder}
\end{equation}
where $z_n=z_{f,le}+\frac{n}{20}(z_{i,le}-z_{f,le})$ denotes the redshift at which the transition between steps occurs, with $n \in [0, 20]$ representing the step index at that time, ordered from present to past, and $\mathcal{H}(z_n - z)$ being the Heaviside step function evaluated at $z_n - z$. As in the first model, the EoS parameter remains almost the same as in $\Lambda$CDM, since the DE density remains constant.

For this new generalised case of the $\Lambda_{\rm s}$CDM model, the scale factor evolves as

\begin{widetext}
\begin{equation}
  \begin{aligned}
&a(t)= \begin{cases}
D^{\frac{1}{3}} \sin ^{\frac{2}{3}}\left[\frac{3}{2} \sqrt{\frac{\Lambda_{\mathrm{le}}}{3}} t+C_n\right], & \text { for } t \leq t_{10}, \\
\left[\frac{3}{2}\sqrt{\frac{\kappa^2\rho_m}{3}} (t-t_{11})+D^{1/6} \sin{C_{11}}\right]^2, & t_{10}\le t\le t_{11},
\\
D^{\frac{1}{3}} \sinh ^{\frac{2}{3}}\left[\frac{3}{2}\sqrt{\frac{\Lambda_{\mathrm{le}}}{3}} t+B_n\right], & \text { for } t \geq t_{11},
\end{cases}
\end{aligned}
\end{equation}
where 
\begin{equation}
  \begin{aligned}
&\begin{aligned}
& C_n=\sqrt{\frac{3\Lambda}{40}}\sum^{20-\mathrm{n}}_{i=1}\left(\sqrt{11-i}-\sqrt{10-i}\right)t_{21-i},\\
&B_n=\text{arcsinh}(D^{-1/2}a_{10}^{-3/2})+\sqrt{\frac{3\Lambda}{40}}\sum^{11-n}_{i=0}\left(\sqrt{i}-\sqrt{i+1}\right)t_{19-i},\\
& D=\sinh ^{-2}\left[\frac{3}{2}\sqrt{\frac{\Lambda}{3}} t_0+B_1\right], \\
& a_{10}=D^{-1/2}\left[\frac{3}{2}\sqrt{\frac{\kappa^2\rho_m}{3}} (t_{10}-t_{11})+D^{1/6} \sin^{1/3}\left({\sqrt{\frac{3\Lambda}{40}}t_{11}+C_{11}}\right)\right]^3,
\end{aligned}
\end{aligned}  
\label{consts}
\end{equation}
\end{widetext}
and $n$ takes values corresponding to the discrete steps with $n \in [0, 20 ]$. The following term also satisfies $D=\Omega_{\mathrm{m},0}/\Omega_{\mathrm{d},0}=\kappa^2\rho_{m,0}/\Lambda$, which also determines the age of the Universe today for a
given $\rho_{\mathrm{m},0}$ and $\Lambda$ using Eq. (\ref{consts}).
 In this type of model, as in $\Lambda_{\rm s}$CDM, even though the scale factor is continuous, the Hubble parameter is not, being smaller than in $\Lambda$CDM. Similarly the total EoS parameter, $w$, will also be discontinuous,  showcasing a singularity at the each step jump:
 \begin{equation}
    w_{\textrm{d,le}}=-1-\frac{(1+z)\sum^{20}_{i=1}\delta(z_n-z)}{3 \left[-10+\sum^{20}_{n=1}\mathcal{H}(z_n-z)\right]},
    \label{eosl}
\end{equation}
where $z_n$, as previously introduced,  denotes the redshift at each step jump. The EoS in this case will be -1 across all regimes, except during the step jump, where it diverges due to the Dirac delta functions.

\subsubsection{Odd ladder-like DE (Lo\texorpdfstring{$\Lambda$}{Lambda}CDM) \label{sec2c}}

Similar to the previous case, we introduce a two-parameters extension of the $\Lambda$CDM model, with  an  $M$-step ladder-function. 
The difference lies in the requirement that  $M$ must be odd, ensuring that there is no moment in time when the DE density becomes zero. We denote the initial and final redshifts of the transition as $z_{i,lo}$ and $z_{f,lo}$ respectively. In fact, the EoS parameter for DE reads as in Eq.~(\ref{eosl})
where $N$, $z_{i,le}$ and $z_{f,le}$ are substituted by $M$, $z_{i,lo}$ and $z_{f,lo}$, respectively, in Eq.~(\ref{densityladder}).
Under those assumption, it is easy to show that for a $M=19$ odd case, the scale factor evolves as:
\begin{widetext}
\begin{equation}
  \begin{aligned}
&a(t)= \begin{cases}
D^{\frac{1}{3}} \sin ^{\frac{2}{3}}\left[\frac{3}{2} \sqrt{\frac{\Lambda_{\mathrm{lo}}}{3}} t+C_n\right], & \text { for } t \leq t_{10}, \\
D^{\frac{1}{3}} \sinh ^{\frac{2}{3}}\left[\frac{3}{2}\sqrt{\frac{\Lambda_{\mathrm{lo}}}{3}} t+B_n\right], & \text { for } t \geq t_{10},
\end{cases}
\end{aligned}
\end{equation}
where
\begin{equation}
  \begin{aligned}
&\begin{aligned}
& C_n=\sqrt{\frac{3\Lambda}{38}}\sum^{20-\mathrm{n}}_{i=1}\left(\sqrt{11-i}-\sqrt{10-i}\right)t_{20-i},\\
&B_n=\text{arcsinh}(\sin{C_{10}})+\sqrt{\frac{3\Lambda}{2\mathrm{M}}}\sum^{n}_{i=1}\left(\sqrt{10-i}-\sqrt{11-i}\right)t_{i},\\
& D=\sinh ^{-2}\left[\frac{3}{2}\sqrt{\frac{\Lambda}{3}} t_0+B_1\right],
\end{aligned}
\end{aligned}  
\end{equation}
\end{widetext}
where $n$ takes values corresponding to the discrete steps with $n\in [0,19]$ in this case. 
As in the first model, the EoS parameter remains \mbox{-1} across nearly all redshifts, except during the transition, where it follows the form given in Eq.~(\ref{eosl}), subject to the slight modification previously described. Similarly to the previous model, 
$n$ denotes the step index, starting from the one closest to the present time and progressing towards earlier epochs. The scale factor has been normalised such that $a(t_0)=1$
at the present time.

\subsection{SSCDM \label{sec2d}}

This model introduces a negative  DE density in the past, which begins a smooth transition at a chosen redshift, eventually reaching a final positive and constant value following a sign change. The energy density of DE in this framework is given by:
\begin{equation}
  \begin{aligned}
&  \Lambda_\mathrm{ss}(x)=\Lambda \begin{cases}-1, & x \le x_{i,ss},\\  1-2(126 t^5 - 420 t^6 \\
+ 540 t^7 - 315 t^8 + 70 t^9),  & x_{i,ss}<x<x_{f,ss},\\
1, & x\ge x_{f,ss},\end{cases} \\
\end{aligned}  
\label{dnesitysscdm}
\end{equation}
where $t=\frac{x-x_{f,ss}}{x_{i,ss}-x_{f,ss}}$ and $x=\ln(a/a_0)$. The model includes two additional parameters, $x_{i,ss} = -\ln(1 + z_{i,ss})$ and $x_{f,ss} = -\ln(1 + z_{f,ss})$, which represent the beginning and end of the transition period and govern its smoothness. This setup is a two-parameter extension of the $\Lambda$CDM approach.

In this type of model, the EoS parameter will differ from that of $\Lambda$CDM, as the DE density is no longer constant during part of the evolution. In fact, the EoS parameter evolves as  
\begin{equation}
  \begin{aligned}
&  w_\mathrm{d,ss}(x)= \begin{cases}-1, & x \le x_{i,ss},\\-1  -\frac{1260}{\Delta x \text{ }\Lambda_\mathrm{ss}(x)/\Lambda}(t^4 \\-4t^5+6t^6-4t^7+t^8),  & x_{i,ss}<x<x_{f,ss},\\
,-1 & x\ge x_{f,ss},\end{cases}\\
\end{aligned}  
\label{eossscdm}
\end{equation}
where $\Delta x = x_{i,ss}-x_{f,ss}$. In models (A) and (B), the EoS parameter exhibits a delta function at each step discontinuity, arising from the derivative of the step function.
 This is no longer the case due to our new model exhibiting a smoother transition. Nevertheless, as in all DE models involving a transition from negative to positive DE density, the EoS parameter $w_{\mathrm{d,ss}}(x)$ diverges at the sign-switching point, undergoing a phantom crossing through infinite values.
 Even though $w_\mathrm{d,ss}(x)$ is not well defined throughout the entire regime $x \in (x_{i,ss}, x_{f,ss})$, this is not the case for the total EoS parameter, as defined in Eq.~(\ref{eq:totalEoS}).

\subsection{ECDM \label{sec2e}}
For this model, we introduce a DE density that evolves according to the interpolating error function. More precisely, we assume
\begin{equation}
    \Lambda_\mathrm{e}(x)=\Lambda \text{ Erf}[\eta(x-x_{\dagger,e})],
    \label{densityecdm}
\end{equation}
where $\eta$ is a parameter that determines the smoothness of the transition, and $x_{\dagger,e} = -\ln(1 + z_{\dagger,e})$ denotes the redshift at which the density changes sign. This setup is also a two-parameter extension of the $\Lambda$CDM approach.

Regarding the EoS parameter, it can be readily obtained by imposing the conservation of the energy momentum tensor of DE:
\begin{equation}
    w_\mathrm{d,e}(x)=-1-\frac{2\eta e^{-\eta^2(x-x_{\dagger,e})^2}}{3\sqrt{\pi}\text{ Erf}[\eta(x-x_{\dagger,e})]}.
    \label{eosecdm}
\end{equation}

As in model (C), the smooth transition exhibited by the DE density in this model eliminates the delta functions that appear in EoS state  in models (A) and (B). However, as in the previous model, since the error function vanishes at the sign-switching point $x = x_{\dagger,e}$, the EoS parameter diverges, undergoing a phantom crossing through infinite values.
 Regarding the total EoS parameter defined in Eq.~(\ref{eq:totalEoS}), there is no problem, showcasing a well defined background value. 

\subsection{Comparing all these models \label{sec2f}}

A few words may be in order to compare the models among each other and in particular with  $\Lambda$CDM. For early-time measurements, the models are similar; the same holds true once the cosmological constant reaches its maximum constant value, as they all behave like $\Lambda$CDM. However, differences emerge during the transition period and in the smoothness of the change. To compare the models, we need to quantify the parameters that will define them. The procedure we will use is as follows:

Let us begin by considering model (B) with an even number of steps. The redshifts $z_{i,le}$, $z_{i,lo}$, $z_{f,le}$, and $z_{f,lo}$ are defined as follows. We start by specifying the sign-switching redshift, $z_{\dagger,l}$, which we take to be $z_{\dagger,l} \sim 1.7$ \cite{Akarsu:2023mfb}, along with the step size $\Delta z_\mathrm{step} = 0.1$, ensuring that all steps have equal width. For the even case, we choose $N = 20$ for the simulations in order to achieve a smoother transition. Taking the sign-switching redshift as the reference point, we move upwards by $N/2$ steps and downwards by another $N/2$ steps. The initial and final transition redshifts can then be computed using the following expressions: $
z_{i,le} = z_{\dagger,l} + \Delta z_\mathrm{step} \cdot \frac{N}{2}$ and $ z_{f,le} = z_{\dagger,l} - \Delta z_\mathrm{step} \cdot \frac{N}{2}$. The odd case follows a similar procedure, using the same sign-switching redshift $z_{\dagger,l}$ and the same step size $\Delta z_\mathrm{step}$. However, the number of steps, denoted by $M$, differs. To construct a comparable model, we set $M = 19$, which is close to the even case, but avoids the $\rho_{\rm d} = 0$ step. Using one fewer step results in a shift of either the initial or final transition redshift. In our case, we choose to keep the final redshift fixed at $z_{f,lo} = z_{f,le}$. The corresponding expressions are:
$z_{f,lo} = z_{\dagger,l} - \Delta z_\mathrm{step} \cdot \frac{M + 1}{2}$ and $z_{i,lo} = z_{\dagger,l} + \Delta z_\mathrm{step} \cdot \frac{M - 1}{2}$
which results in a reduced initial redshift. This construction facilitates a direct comparison between the even and odd step models. 

Regarding these same parameters for model (C), the procedure was different, as they were set to mimic the shape of the DE density of model (B) with even steps, thereby ensuring a similar transition speed and a sign-switch at a comparable redshift, $z_{\dagger\mathrm{,ss}}\sim1.72$. The transition speed of model (D), $\eta$, has been set to behave similarly to models (B) and (C) and to the already studied model $\Lambda_\mathrm{s}$VCDM \cite{Akarsu:2024eoo}. For all of the simulations in this work, we have set the free parameters of each model with the same values \footnote{For all simulations we adopt the value $H_0 = 69.68\,\mathrm{km/s/Mpc}$, as obtained for the $\Lambda_\mathrm{s}$CDM model using the Planck2018+Pantheon data in Ref.~\cite{Akarsu:2022typ}.}. For model (A) $z_{\dagger,s}=1.7$; model (B) $N=20$, $z_{i,le}=2.7$, $z_{f,le}=0.7$, $M=19$, $z_{i,lo}=2.6$, $z_{f,lo}=0.7$; model (C) $z_{i,ss}=2.7$, $z_{f,ss}=1$ and model (D) $\eta=10$ and $z_{\dagger,e}=1.7$. 

\begin{figure}[t]
    \centering
    \includegraphics[width=1.0\linewidth]{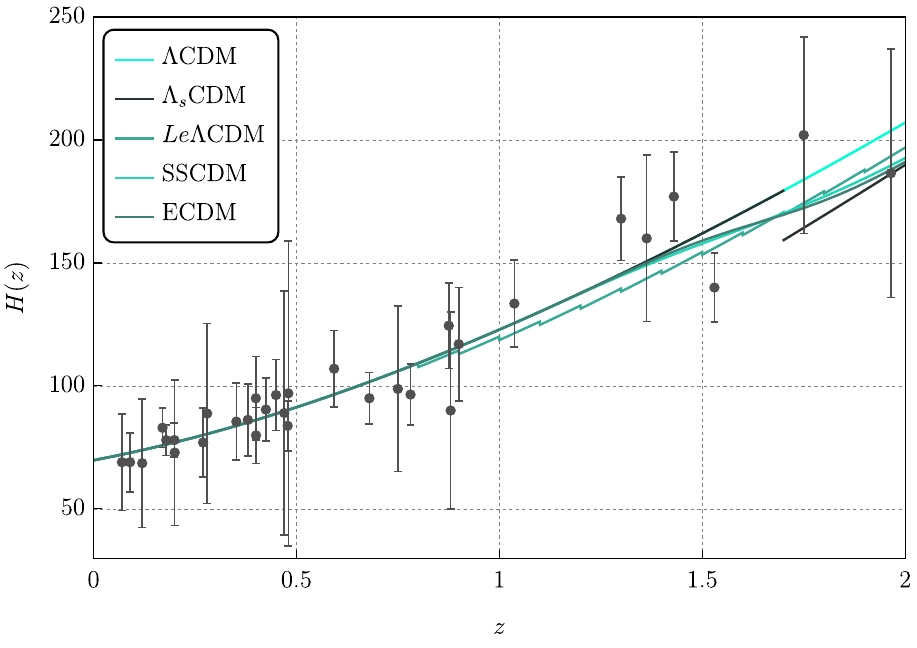}
    \caption{\justifying{\it{$H(z)$ in units of $Km/s/Mpc$ against redshift. The Hubble constant was set from Ref.~\cite{Akarsu:2022typ} and the error bars were obtained from Ref.~\cite{Favale:2023lnp}.}}}
    \label{fig:hubble}
\end{figure}

We can classify the models into two groups. On the one hand, we have models (A) and (B), which possess a discontinuous DE density. This means that, even though the scale factor is continuous, the Hubble parameter remains discontinuous. These models, in what refers to the EoS of DE, behave more similarly to the concordance cosmology, as the EoS parameter remains \mbox{nearly -1} throughout most of the cosmological evolution.  On the other hand, models (C) and (D) feature a DE density that varies over time during the sign-switching process.  However, for these two models the continuity of the DE density ensures both a continuous Hubble parameter and scale factor evolution\footnote{Unlike models (A) and (B), we were unable to obtain an analytical expression for the cosmic evolution of the scale factor in models (C) and (D).}. Models (C) and (D)  exhibit a behaviour that differs more significantly from the standard model, as the EoS evolves over time. This variation will be more evident in the cosmographic analysis to be conducted in the following section.

\begin{figure}[t]
    \centering
    \includegraphics[width=1.0\linewidth]{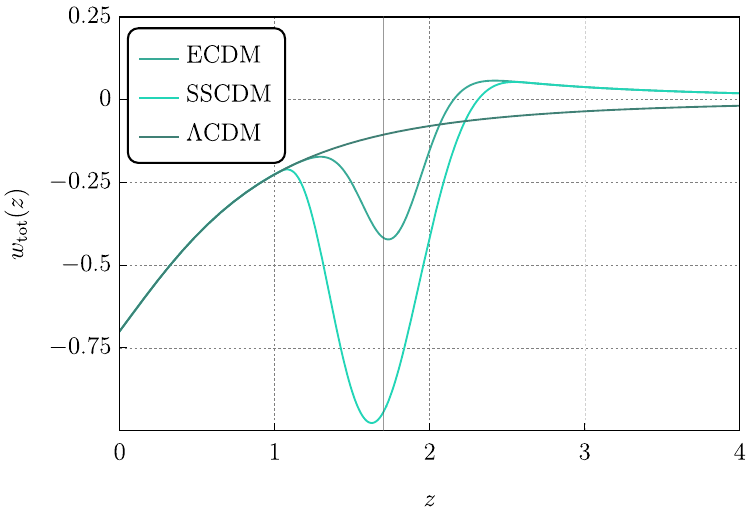}
    \caption{\justifying{\it{Total EoS parameter, $w_{\mathrm{tot}}(z)$, against redshift. The vertical black line denotes the sign-switching redshift for the DE density. Models (A) and (B) exhibit behaviour similar to  $\Lambda$CDM.}}}
    \label{fig:wtot}
\end{figure}

Upon observing Fig. \ref{fig:hubble}, we note that the Hubble parameter, $H(z)$, is discontinuous for models (A) and (B), which will lead to further discontinuities in subsequent time derivative of the Hubble rate or quantities related to it. Our models fit\footnote{This is an approximated ``fit'' based on the values obtained in \cite{Akarsu:2025ijk,Escamilla:2025imi}, on the sense that we took the values of $H_0$, $z_\dagger$, $\eta$, $\Omega_{\mathrm{m},0}$, $\Omega_{\mathrm{r},0}$ and $\Lambda$ and adjusted them to our model. We are currently observationally fitting properly these models and we hope to report soon our results.} most of the cosmic chronometers obtained from Ref. \cite{Favale:2023lnp}, with a particular improvement for the oldest value at $z=1.975$.

Regarding the EoS parameter, since the energy density remains constant at all times in models (A) and (B), except for the step jumps, these models almost behave like perfect fluids with an EoS parameter of $w_d=-1$, except for the divergence at the sign-switching for model (A) and the divergences at each step jump for model (B). The same is not true for models (C) and (D), where $w_d$ diverges near the sign-switching point. In both cases, prior to the transition $w_d>-1$, tending towards +$\infty$. It then undergoes a bounce through infinity, and after the sign-switching, reappears from $-\infty$, remaining at $w_d<-1$ before it converges to $w_d=-1$. Nonetheless, the total EoS parameter, given by Eq. (\ref{eq:totalEoS}), is well-defined, as it can be seen in Fig. \ref{fig:wtot}. 

The four models exhibit a singularity at the sign-switching time. As it can be seen in Ref. \cite{Paraskevas:2024ytz}, model (A) has already been proven to have a type II singularity and the same happens with model (B) due to its step-like structure, nonetheless, this last model also exhibits singularities at each of the step jumps. These kind of singularities have already been studied in the existing literature \cite{Barrow1,Barrow:2004xh,Nojiri:2005sx,Gorini:2003wa,Fernandez-Jambrina:2006tkb}, especially in the case of the abrupt transitioning model $\Lambda_{\rm s}$CDM \cite{Paraskevas:2024ytz}, proving that, given that the singularities are relatively weak, they alone can not dissociate any bound system before the continual expansion of the universe does. Thus, the dissociation of bound orbits is primarily driven by cosmic expansion,
with the singularity inducing a minor perturbation that
slightly increases the Hubble value at a specific instant.
This perturbation contributes minimally to the dissociation of a bound orbit. Regarding models (C) and (D), they do exhibit a type II singularity in the limit case where the transition is made to be instantaneous as to mimic model (A), as proven in Appendix \ref{sing}. Nevertheless, when the transition is smooth a type V singularity arises, also known as a $w$ singularity \cite{Dabrowski:2009kg,Bouhmadi-Lopez:2019zvz}, due to the divergence of the DE EoS parameter, whilst the scale factor, density and pressure remain finite. A more profound study of said singularities will be carried out in future work to study its impact in the dynamics and evolution of bounded structure.
 
\section{Cosmographic analysis}\label{sec3}

We can use a cosmographic analysis to compare the  aforementioned models by studying their cosmographic parameters \cite{Visser:2004bf,Cattoen:2007sk,Capozziello:2008qc,Capozziello:2008qc}. For such approaches the scale factor is Taylor expanded around the present day value $a_0$ as
\begin{equation}
    a(t)=a_0\left[1+\sum_{n=1}^\infty\frac{A_n(t_0)}{n!}[H_0(t-t_0)]^n\right],
\end{equation}
where the cosmographic parameters $A_n$ are defined as 
\begin{equation}
    A_n:=a^{(n)}/(aH^n) \quad ,n\text{ }\in\text{ }\mathbb{N},
    \label{an}
\end{equation}
$a^{(n)}$ being the $n^{th}$ derivative of the scale factor with respect to cosmic time. Some of these  parameters are also known as, the deceleration parameter $q=-A_2$, the jerk $j=A_3$, the snap $s=A_4$ and the lerk $l=A_5$. Based on the cosmographic approach, the statefinder hierarchy was developed to distinguish various DE models. These expressions allow us to define the \textit{Statefinder hierarchy} \cite{Albarran:2017kzf},
\begin{equation}
    \begin{aligned}
        & S_3^{(1)}=A_3, \\
        & S_4^{(1)}=A_4+3(1-A_2), \\
        & S_5^{(1)}=A_5-2(4-3A_2)(1-A_2).
    \end{aligned}
\end{equation}
For concordance cosmology or $\Lambda$CDM model

\begin{equation}
    S_n^{(1)}|_{\Lambda CDM}=1.
    \label{sn1}
\end{equation}

A \textit{null diagnostic} is defined for concordance cosmology by Eq. (\ref{sn1}), since some of the equalities are likely to be violated by evolving DE models. Together with these  parameters, we also introduce the following statefinder parameter \cite{Sahni:2002fz}

\begin{equation}
    \tilde{s}=\frac{1-S_3^{(1)}}{3(A_2+\frac{1}{2})}.
\end{equation}

We have carried out the cosmographic analysis in two parts, first studying the cosmographic parameters and then carrying out a parametrisation of the statefinder hierarchy. We next present our results.

\begin{figure*}[htbp]
    \centering
    \begin{subfigure}{0.4\textwidth}
        \centering
        \includegraphics[width=\textwidth]{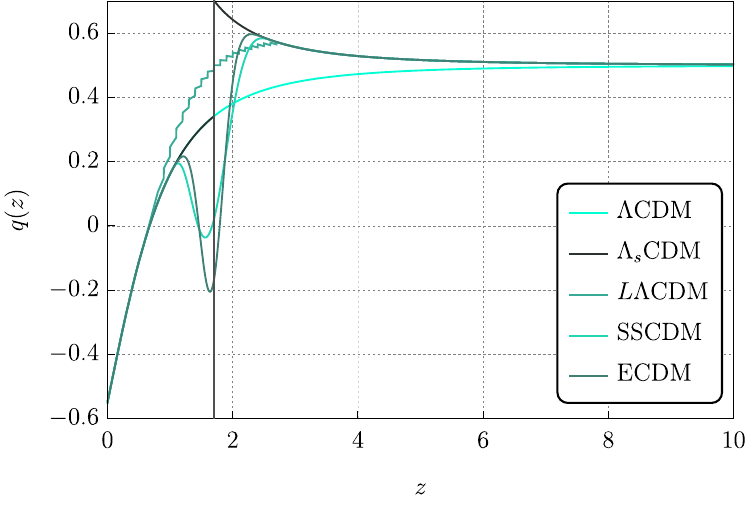}
        \label{fig:qz}
    \end{subfigure}
    \hfill
    \begin{subfigure}{0.4\textwidth}
        \centering
        \includegraphics[width=\textwidth]{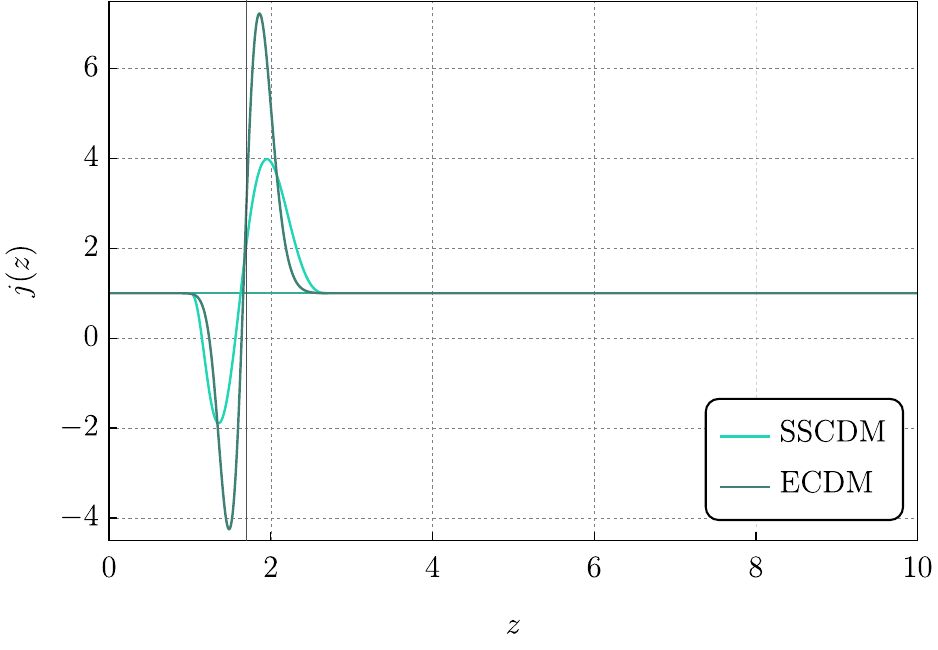}
        \label{fig:jz}
    \end{subfigure}
    
    \begin{subfigure}{0.4\textwidth}
        \centering
        \includegraphics[width=\textwidth]{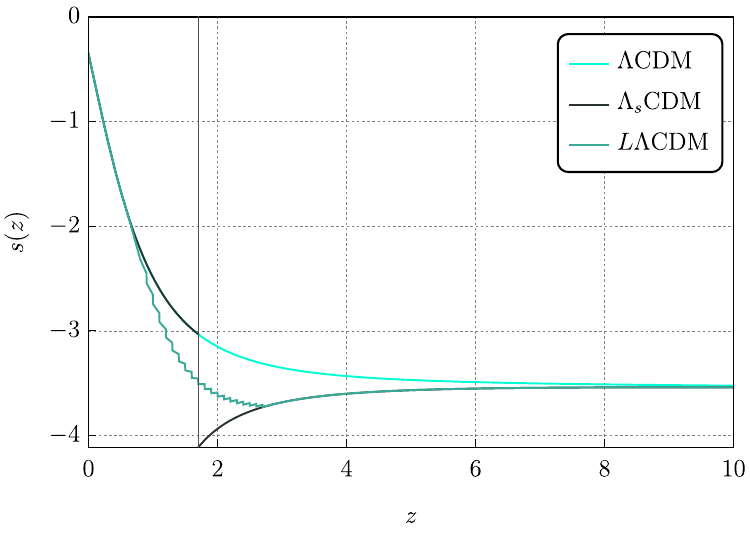}
        \label{fig:jz_model1}
    \end{subfigure}
    \hfill
    \begin{subfigure}{0.41\textwidth}
        \centering
        \includegraphics[width=\textwidth]{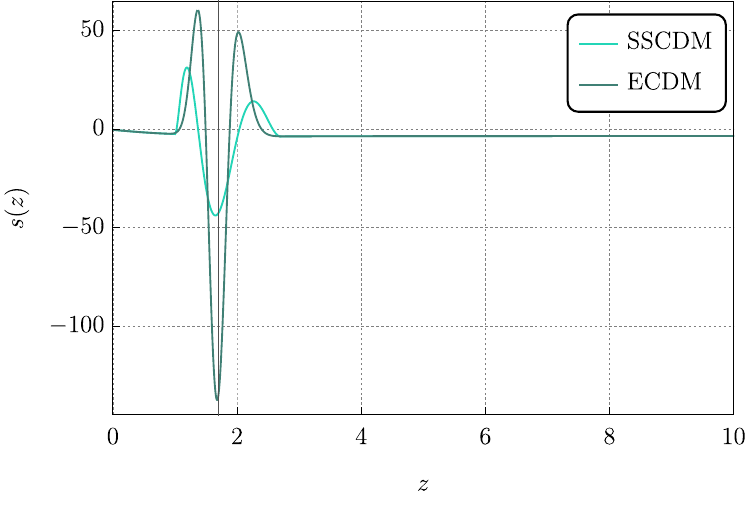}
        \label{fig:jz_model2}
    \end{subfigure}

    \begin{subfigure}{0.4\textwidth}
        \centering
        \includegraphics[width=\textwidth]{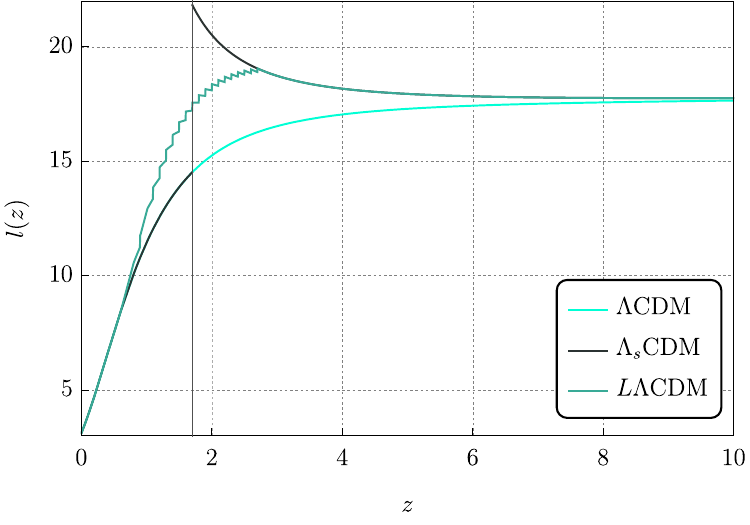}
        \label{fig:lz_model1}
    \end{subfigure}
    \hfill
    \begin{subfigure}{0.42\textwidth}
        \centering
        \includegraphics[width=\textwidth]{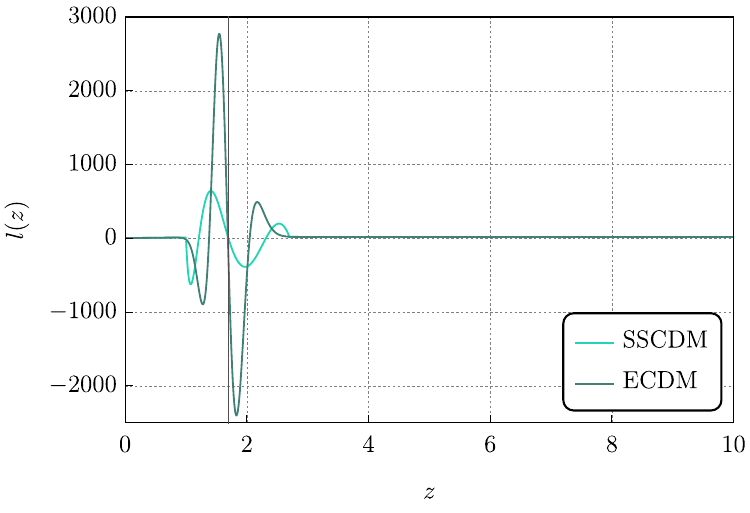}
        \label{fig:lz_model2}
    \end{subfigure}
  
    \caption{\justifying{\it{Comparison of the cosmographic parameters $q(z)$, $j(z)$, $s(z)$ and $l(z)$ against redshift for different models (as stated on the plots and in the main text of this subsection). In this figures we only considered the even step (N=20) ladder model. The top left image showcases the deceleration parameter for all of the models whilst the top right image showcases the jerk parameter. In the middle row, the snap parameter is shown for the different models. The plots have been grouped according to the scale taken by the snap parameter. The lerk parameter can be observed in the last row with scale dependent plots as well.  The assumed values for the current energy densities are $\Omega_{\mathrm{m},0}=0.3$, $\Omega_{\mathrm{r},0}=0$ and $\Omega_{\mathrm{d},0}=0.7$. The vertical darker line indicates the redshift at which the DE density changes sign. }}}
    \label{fig:all_plots}
\end{figure*}

\subsection{Cosmographic parameters}

In this section, we examine the evolution of the cosmographic parameters at low redshifts. The radiation energy density is set to zero, $\Omega_{\mathrm{r},0}=0$, as we focus solely on low-redshift regimes around the time of the sign-switching event and thereafter.

\begin{figure*}[htbp]
    \centering
    \begin{subfigure}{0.45\textwidth}
        \centering
        \includegraphics[width=\textwidth]{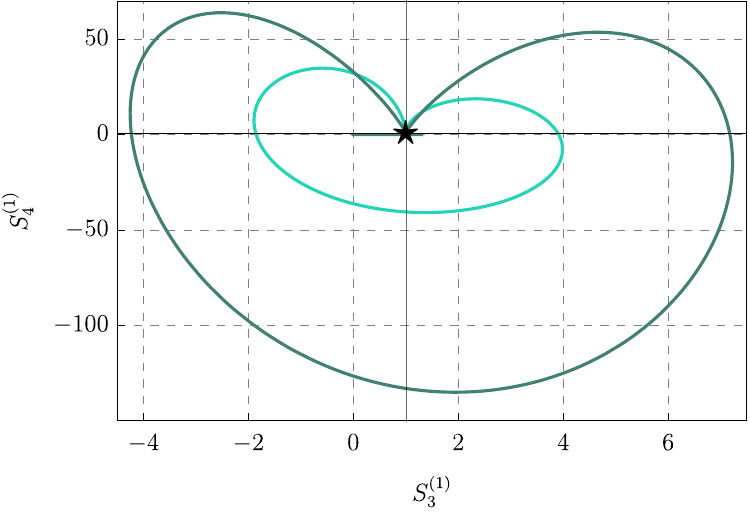}
    \end{subfigure}
    \begin{subfigure}{0.45\textwidth}
        \centering
        \includegraphics[width=\textwidth]{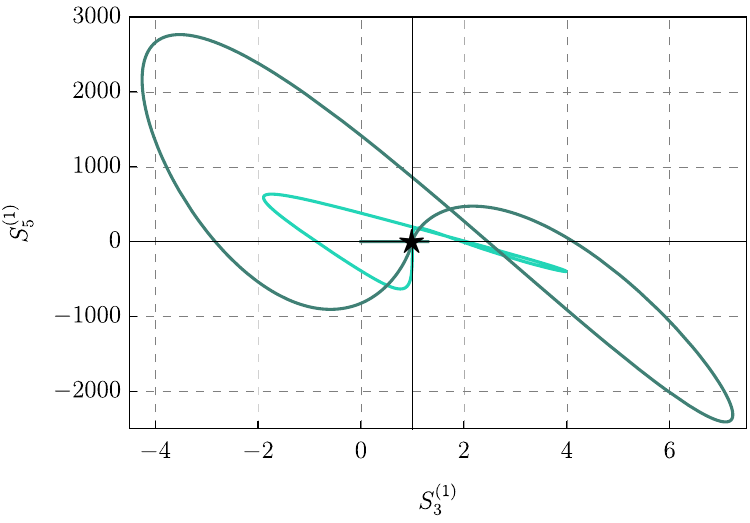}
    \end{subfigure}
    \caption{\justifying{\it{In these plots $\{S_3^{(1)},S_4^{(1)}\}$ (left) and $\{S_3^{(1)},S_5^{(1)}\}$ (right) are depicted. We have assumed a vanishing amount of radiation. The models corresponding to $\Lambda$CDM, $\Lambda_{\rm s}$CDM , and both ladder models assume constant values of \mbox{$S_3^{(1)}=S_4^{(1)}=S_5^{(1)}=1$.} The lighter green line represents the SSCDM model, or model (C), while the darker green line represents the ECDM model, or model (D). Indeed,  the models have been identified using the same colours as in Fig. \ref{fig:all_plots}. The assumed  values of the actual energy densities are $\Omega_{\mathrm{m},0}=0.3$, $\Omega_{\mathrm{r},0}=0$ and $\Omega_{\mathrm{d},0}=0.7$. The stars indicate the present day values of the statefinder parameters for each of the models. As can be seen, they all overlap in the same region.}}}
    \label{fig:statefinde2r}
\end{figure*}

In the top left image of \mbox{Fig.~\ref{fig:all_plots}}, we can observe the evolution of the deceleration parameters, $q(z)$, for the four models and $\Lambda$CDM. Models (A) and (B) behave as expected, entering into an accelerated expansion era around the same time as $\Lambda$CDM. In the case of model (B), only the even case can be observed in Fig. \ref{fig:all_plots}, as both the odd and even cases showcase similar behaviours, the only difference being the number of steps. Nonetheless, the models in which the DE density varies exhibit some degrees of fluctuation. Models (C) and (D) decrease the deceleration rate faster than the previously mentioned models, with the deceleration parameter reaching a local negative minimum after entering an acceleration phase. Subsequently, as the deceleration parameter becomes positive again, the Universe temporarily re-enters a decelerating epoch before $q$ turns negative once more, initiating the current accelerated phase of the Universe and reaching $q(z) \sim -0.55$, similar to the value in $\Lambda$CDM.

In the top right image of Fig.~\ref{fig:all_plots}, we present the evolution of the jerk parameter, $j(z)$. For all models with a constant DE density, the jerk remains fixed at $j(z) = 1$. This behavior is characteristic of $\Lambda$CDM models, where $j = 1$ holds consistently, provided that radiation is neglected. In contrast, models with a time varying DE density exhibit some oscillations in $j(z)$ around the epoch of sign switching. Nonetheless, away from this transition period, the jerk parameter returns to unity, as expected.

In the middle row of Fig.~\ref{fig:all_plots}, we present the evolution of the snap parameter, $s(z)$, categorising the models according to the scale at which $s(z)$ evolves. The left panel displays models with a constant DE density, which as expected, converge to the $\Lambda$CDM model.  In the right-hand figure, models (C) and (D) exhibit oscillations in $s(z)$ around the epoch of sign switching. As the DE density approaches the expected $\Lambda$CDM value, the snap parameter begins to increase, in a manner consistent with models (A) and (B).

Finally, on the last row of Fig.~\ref{fig:all_plots} we have plotted the evolution of the lerk parameter, $l(z)$.  As in the previous cases, on the left image we observe the convergence of the constant DE density models.
In the right figure some oscillations can be observed for both models, which deviate from the standard cosmological model. The oscillations of the cosmographic parameters in both models (C) and (D) could serve as a straightforward way to rapidly assess the viability of these models. The absence of such oscillations—should we ever be able to measure the relevant parameters would readily discredit these models.

\subsection{Statefinder hierarchy}

\begin{figure*}[htbp]
    \centering
    \begin{subfigure}{0.3\textwidth}
        \centering
        \includegraphics[width=\textwidth]{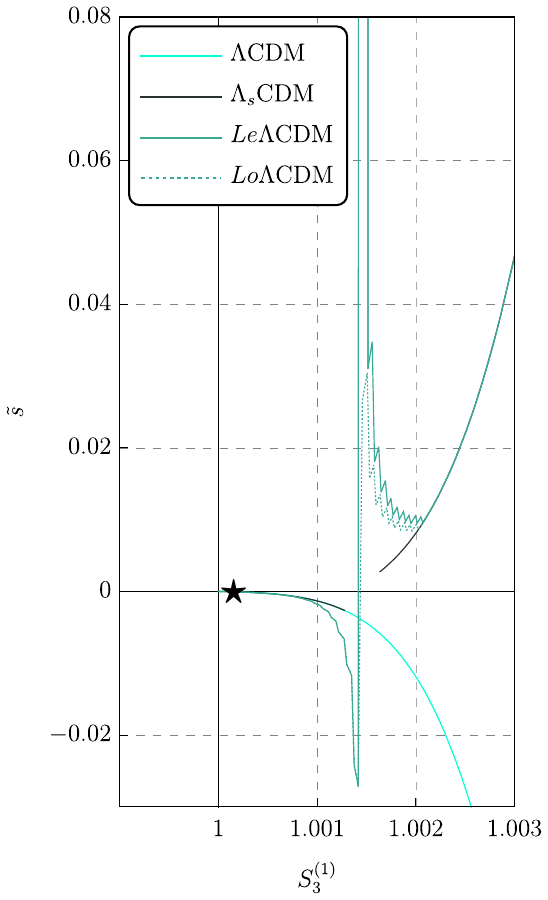}
    \end{subfigure}
    \begin{subfigure}{0.3\textwidth}
        \centering
        \includegraphics[width=\textwidth]{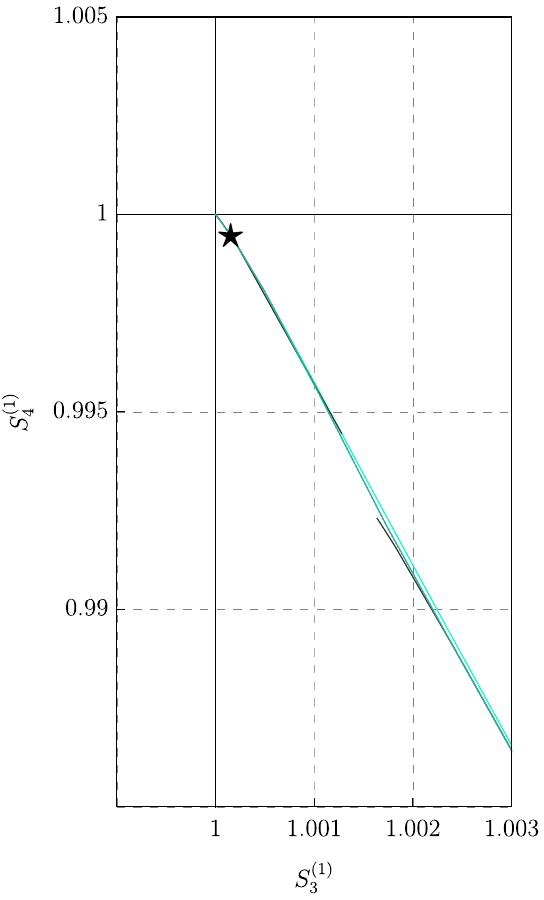}
    \end{subfigure}
    \begin{subfigure}{0.3\textwidth}
        \centering
        \includegraphics[width=\textwidth]{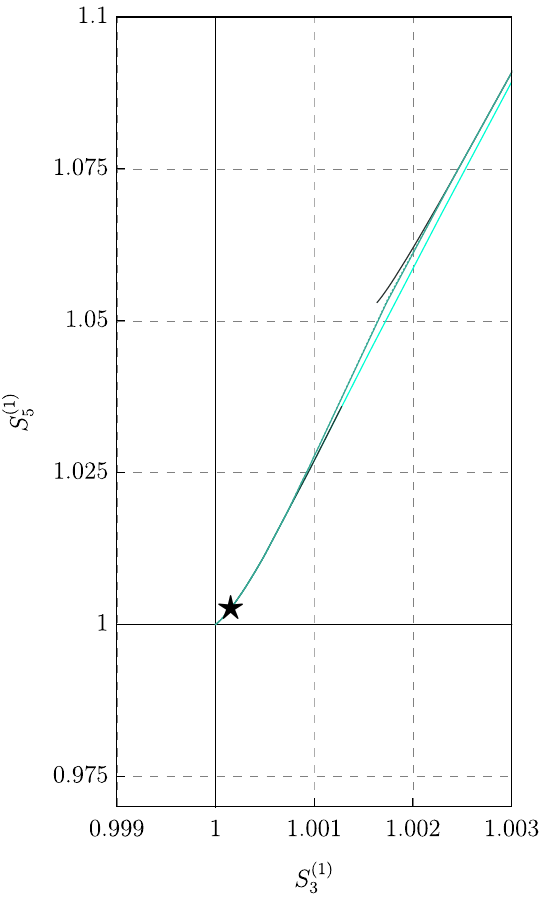}
    \end{subfigure}
    \caption{\justifying{\it{In these plots $\{S_3^{(1)},s\}$ (left), $\{S_3^{(1)},S_4^{(1)}\}$ (middle) and $\{S_3^{(1)},S_5^{(1)}\}$ (right) are depicted for $\Lambda$CDM, $\Lambda_{\rm s}$CDM, and both ladder models. We have assumed a very small, yet non-vanishing, amount of radiation.
    Therefore, the statefinder parameters $S_3^{(1)}$, $S_4^{(1)}$, $S_5^{(1)}$ and $\tilde{s}$ deviate from unity and zero, respectively.
    The stars indicate the present day values of the statefinder parameters for each of the models, which in all four cases are overlapping. The models have been plotted using the same colours as in Fig. \ref{fig:all_plots}.  The assumed  values of the actual energy densities are
    $\Omega_{\mathrm{m},0}=0.3$, $\Omega_{\mathrm{r},0}=8\times 10^{-5}$ and $\Omega_{\mathrm{d},0}=1-\Omega_{\mathrm{m},0}-\Omega_{\mathrm{r},0}$.}}}
    \label{radiation}
\end{figure*}

In \mbox{Fig.~\ref{fig:statefinde2r}}, we present the late-time evolution of the statefinder parameters: $\{S_3^{(1)},S_4^{(1)}\}$ (left) and $\{S_3^{(1)},S_5^{(1)}\}$ (right), for the four models analysed on  this paper, in addition to $\Lambda$CDM. We have assumed, as a starting ansatz, a vanishing radiation energy density, thereby simplifying the parametrisation of the models. Moreover, the statefinder parameters of model (A) and both ladder models coincide with those of $\Lambda$CDM. In fact, the statefinder parameters are specifically designed to remain constant in the $\Lambda$CDM scenario when radiation is neglected. Specifically, for $\Lambda$CDM, the statefinder parameters $S_3^{(1)}$, $S_4^{(1)}$, $S_5^{(1)}$, and $\tilde{s}$ take values of unity and zero, respectively. Therefore, this approach offers a model-independent means of distinguishing evolving DE models from the cosmological constant.

Although the statefinder parametrisation is not particularly effective at distinguishing models with a constant DE component; indeed, the standard cosmographical parametrisation performs significantly better, as shown in Fig.~\ref{fig:all_plots}; this statement no longer holds true when considering dynamical DE models that vary continuously from negative to positive values of their DE density. In this case, Fig.~\ref{fig:statefinde2r} clearly demonstrates that the two models, i.e. constant and evolving DE  models, can be distinguished.

In order to compare our model (B); i.e. L$\Lambda$CDM, with even and odd steps, to the $\Lambda$CDM and $\Lambda_{\rm s}$CDM models, we will also include the radiation component (although it is almost negligible today). The reason for including radiation is that the statefinder parameters deviate from their constant value for the $\Lambda$CDM, $\Lambda_{\rm s}$CDM , and ladder models. Therefore, by looking back into the past, we may be able to distinguish between them, at least from a theoretical point of view.

For the remainder of this subsection, we shall assume the presence of radiation. This could, at least in theory, allow us to distinguish our ladder-like DE model; i.e., the L$\Lambda$CDM setup,  from both $\Lambda$CDM and $\Lambda_{\rm s}$CDM. In Fig.~(\ref{radiation}), the statefinder hierarchy is shown for three parametrisations: $\{S_3^{(1)}, \tilde{s}\}$ (left), $\{S_3^{(1)}, S_4^{(1)}\}$ (middle), and $\{S_3^{(1)}, S_5^{(1)}\}$ (right) and for the four models analysed in presence of radiation. By carefully examining the statefinder hierarchy parametrisation shown in Fig.~(\ref{radiation}), we observe that the four models converge in the past, when DE is no longer dominant. However, some differences are evident during the sign transition of the cosmological constant, particularly in the $\{S_3^{(1)}, \tilde{s}\}$ representation.
In fact, we wish to emphasise that, in the case of the $\{S_3^{(1)}, \tilde{s}\}$ 
parametrisation, a slight difference is observed between the even and odd step ladder models. Specifically, for the odd step ladder model, the cosmological constant never vanishes during the transition process; hence, the parametrisation remains well defined, exhibiting a smoother transition, as one might expect. This is not the case for the even step ladder model, where the cosmological constant vanishes, inducing discontinuities in the $\tilde{s}$ parameter.
The gap observed in the plots for model (A) arises from the model's underlying behaviour.
In the aforementioned figure, the stars indicate the present-day values of the statefinder parameters for each of the models and as can be seen, they all overlap in the same region. It is straightforward to verify that asymptotically (i.e., as $a \rightarrow +\infty$) all models converge to the same point, \mbox{$S_3^{(1)} = S_4^{(1)} = S_5^{(1)} = 1$} and $\tilde{s} = 0$. These are precisely the values attained by the concordance $\Lambda$CDM model at any moment in time in the absence of radiation.

\section{Conclusions \label{sec5}}

Motivated by the compelling $\Lambda_{\rm s}$CDM model, which for brevity we shall refer to as model (A), we introduce and examine a novel dark energy (DE) framework based on a generalised ladder step energy density function, hereafter referred to as model (B) or L{$\Lambda$}CDM. This model is designed to capture potential abrupt transitions in the DE density throughout cosmic time. Actually, it can be divided into two cases, depending on whether the cosmological constant is zero or non-zero when transitioning from negative to positive values. We also analyse two alternative scenarios in which the cosmological constant undergoes a sign change at a specific redshift. We refer to these models as models (C) and (D), or SSCDM and ECDM, respectively. In these latter scenarios, the transitions are implemented smoothly and continuously, allowing us to explore the implications of such dynamical behaviour in a physically consistent manner. In summary, this work introduces several new models featuring a sign-switching DE density that can vary over time, and compares them with the previously proposed $\Lambda_{\rm s}$CDM model. One of the primary aims of this study is to approach as closely as possible a realistic realisation of the sign-switching behaviour of the cosmological constant, or of an evolving DE density. We undertake a detailed and careful analysis of the background evolution of these models, presenting, wherever possible, the analytical time evolution of the scale factor. 

We have also conducted a cosmographic study of the aforementioned models, computing the cosmographic parameters (\ref{an}). The cosmological evolution of these parameters for all the models analysed in this paper is presented in Fig.~\ref{fig:all_plots}. In said figure, one can observe that, in contrast to the more abrupt behaviour of models (A) and (B), models (C) and (D) exhibit a continuous and smooth deceleration parameter with a minimum, distinguishing them from the other models. Regarding the snap, lerk, and jerk parameters: these parameters are discontinuous in models (A) and (B), and continuous in models (C) and (D).

We have completed our research by conducting an analysis of the statefinder hierarchy parameters.  
In our analysis, we have started neglecting radiation. Although in this case, the statefinder parametrisation is not particularly effective at distinguishing models with a constant DE component (Models (A) and (B)) as illustrated in Fig.~\ref{fig:statefinde2r}, the same is not true for evolving DE models (Models (C) and (D)) (cf. also Fig.~\ref{fig:statefinde2r}). As the statefinder parameters were originally designed to remain constant over time for the concordance $\Lambda$CDM setup; or any model with a constant cosmological constant, whether positive or negative, plus dust; we have extended our analysis to include radiation, thereby rendering the statefinder parameters to no longer remain constant for the concordance model. As shown in Fig.~(\ref{radiation}), the models (A), (B), and $\Lambda$CDM can be distinguished around the point at which the cosmological constant changes sign, once radiation is taken into account. Indeed, certain differences become apparent during the sign transition of the cosmological constant. In particular, discontinuities of the statefinder parameters happen in the region where the cosmological constant changes sign for any of the models (A) and (B). This is apparent in Fig.~\ref{radiation} as well as in the analytical expressions presented in Appendix \ref{appendixd}.

Although the differences between the models discussed here are primarily reflected in their cosmographic parameters, several avenues exist to assess whether such differences could be detected observationally. In principle, cosmological fits — which we are currently performing — can indicate whether observational data favour any specific model over others. Furthermore, some of the cosmographic parameters computed in this work can, in fact, be constrained with present observational tools, as demonstrated in Ref. \cite{Luongo:2024fww}, albeit with some degree of model dependence. Model-independent approaches, such as those explored in Ref. \cite{Velazquez:2024aya}, may provide complementary constraints. In addition, perturbative analyses, e.g. using redshift-space distortion data to obtain $f\sigma_8$ measurements as in Ref.\cite{pert}, offer another route to distinguish these models from a standard cosmological constant. Finally, the study of the singularities predicted by these scenarios — already carried out for the $\Lambda_{\rm s}$CDM case in Ref. \cite{Paraskevas:2024ytz} — may yield further observable signatures, particularly regarding their impact on bound cosmic structures. We plan to pursue these analyses in future work to evaluate the viability of the proposed models and to determine whether they can be observationally discriminated.

One must go beyond the background analysis to study structure formation. To this end, perturbation theory provides a mathematical framework for analysing the inhomogeneities of the Universe and deducing observable quantities, such as the distribution of $f\sigma_8$, which enables us to compare the expected results of models with observations of matter clustering. For this reason, in a subsequent paper, we will examine the evolution of cosmological perturbations \cite{pert}. A proper fitting of these models using currently available data will also be addressed in a forthcoming paper \cite{fit}.

\section*{Acknowledgements}

The authors are grateful to \"{O}zg\"{u}r Akarsu,  Carlos G. Boiza, Hsu-Wen Chiang, Nihan Kat{\i}rc{\i} and Thomas Broadhurst for discussions and insights on the current project. 
M. B.-L. is supported by the Basque Foundation of Science Ikerbasque. Our work is supported by the Spanish Grant PID2023-149016NB-I00 funded by (MCIN/AEI/10.13039/501100011033 and by “ERDF A way of making Europe). This work is also supported by the Basque government Grant No. IT1628-22 (Spain) and in particular B. I. U. is funded through that Grant.

\appendix

\section{DE singularities in sign switching models} \label{sing}

In this paper, we have analysed several models that, either through a smooth or sudden transition in the sign of the cosmological constant, lead to some kind of singularity. The first of these  singularities is a Type II (sudden) singularity \cite{Nojiri:2005sx} that appears in  Model (A) at the transition point:
\begin{equation}
    \begin{aligned}
    t=t_\dagger, &\quad a=a_\dagger<\infty, \\ \rho=\rho(a_\dagger)<\infty, &\quad |P_{\textrm{tot}}(a_\dagger)|\rightarrow\infty
    \end{aligned}
\end{equation}
as was proven in Ref. \cite{Paraskevas:2024ytz}. This result can be readily extrapolated to our ladder models; i.e. model (B), where there are sudden singularities at each step jump. 

We next examine the type of singularity that could arise in the other two models. For generality, our analysis will also include the radiation component, although our stated results do not depend on the presence or absence of radiation in the models.

\subsection{SSCDM singularities}

Using Eq. (\ref{eossscdm}), we can compute the total pressure and the total density
\begin{equation}
\begin{aligned}
    & \rho_{\textrm{tot}}= \rho_{\textrm{r},0} e^{-4x}+ \rho_{\textrm{m},0} e^{-3x}+ \rho_{\textrm{d},0} \Lambda_\mathrm{ss}(x), \\
    & P_{\textrm{tot}}=\frac{1}{3}\rho_{\textrm{r},0} e^{-4x}+ w_{\mathrm{d,ss}}(x)\rho_{\mathrm{d},0} \Lambda_\mathrm{ss}(x).
\end{aligned}
\label{sscdmtot}
\end{equation}
Upon evaluating \ $\rho_{\textrm{tot}}(x_{\dagger,ss})$ and $P_{\textrm{tot}}(x_{\dagger,ss})$, we find

\begin{equation}
\begin{aligned}
     \rho_{\textrm{tot}}(x_{\dagger,ss}) & = \rho_{\textrm{r},0} e^{-4x_{\dagger,ss}}+ \rho_{\textrm{m},0} e^{-3x_{\dagger,ss}}, \\
     P_{\textrm{tot}}(x_{\dagger,ss}) & = \frac{1}{3}\rho_{\textrm{r},0} e^{-4x} \\
    & -\rho_{\textrm{d},0}\frac{1260}{\Delta x} (t_{\dagger}^4-4t_{\dagger}^5+6t_{\dagger}^6-4t_{\dagger}^7+t_{\dagger}^8),
    \label{AppendixEq}
\end{aligned}
\end{equation}
where $\Delta x= x_{i,ss}-x_{f,ss}$. Note that $\rho_{\textrm{tot}}(x_\dagger)$ does not depend on $\Delta x$ and remains finite, while $P_{\textrm{tot}}(x_\dagger)$ for the range of redshift considered is negative but finite for finite values of $\Delta x$. Nonetheless, when taking $\Delta x\rightarrow \infty$ the DE density reduces to the abrupt case $\Lambda_\textrm{s}$CDM. For such a limit
 \begin{equation}
     \lim_{\Delta x\rightarrow\infty}P_{\textrm{tot}}(x_{\dagger,ss})=-\infty.
 \end{equation}

 \begin{figure}[t]
    \begin{subfigure}{0.46\textwidth}
        \includegraphics[width=\textwidth]{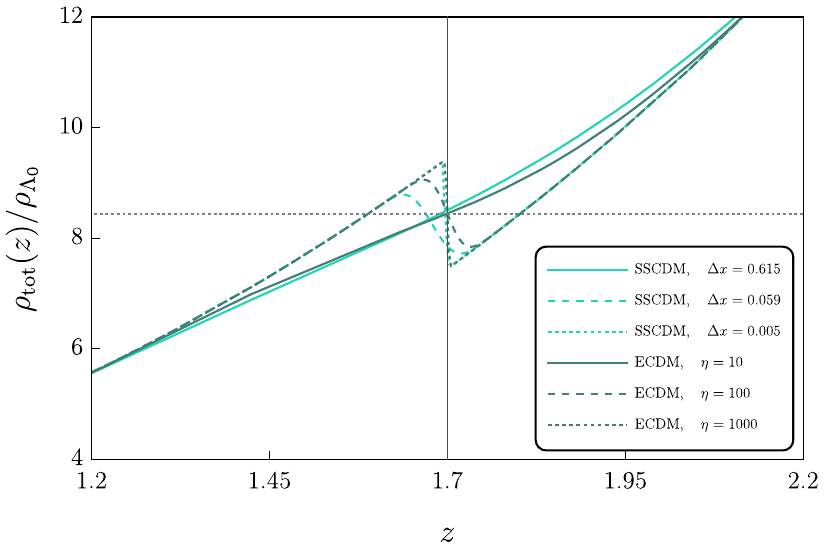}
    \end{subfigure}
    \begin{subfigure}{0.45\textwidth}
        \includegraphics[width=\textwidth]{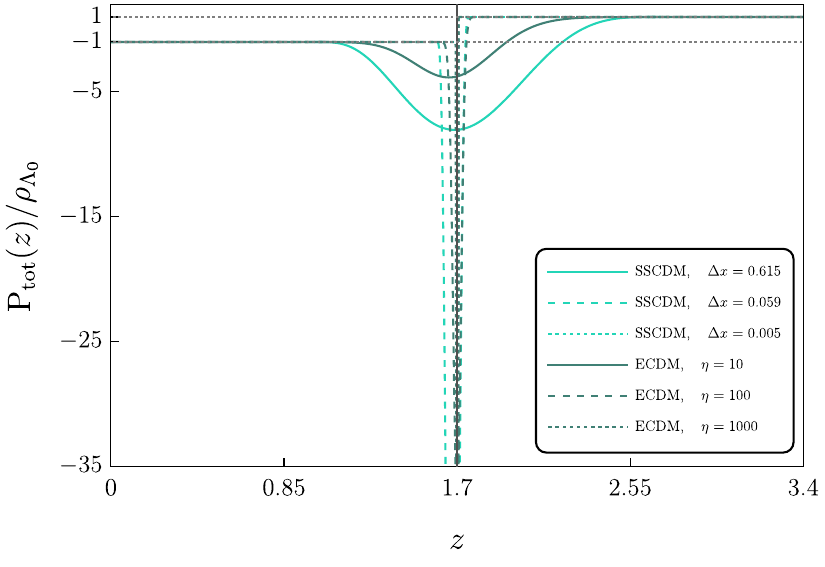}
    \end{subfigure}
    \caption{\justifying{\it{Total density and total pressure of the SSCDM and ECDM universes (see Eqs. (\ref{sscdmtot}) and (\ref{ecdmtot})) with respect to scale factor. For $\Delta x\rightarrow\infty$ and $\eta\rightarrow\infty$,
we observe that $\rho_{\textrm{tot}}(a_\dagger)\rightarrow$ const.$>0$ while P$_{\textrm{tot}}(a_\dagger)\rightarrow-\infty$. The values assumed for the current energy densities are $\Omega_{\mathrm{m},0}=0.3$, $\Omega_{\mathrm{r},0}=8\times 10^{-5}$ and $\Omega_{\mathrm{d},0}=1-\Omega_{\mathrm{m},0}-\Omega_{\mathrm{r},0}$.}}}
    \label{fig:sing}
\end{figure}

This can be easily seen in Fig. (\ref{fig:sing}). Thus, this model at the limit $\Delta x\rightarrow\infty$ is characterised by a type II (sudden) cosmological singularity \cite{Nojiri:2005sx,Barrow:2004xh}. As we previously mentioned the EoS parameter of DE is undefined at the sign-switching time, when the singularity appears, which is a feature of sudden singularities. 

We next analyse what happens for the continuous case, i.e. for a finite $\Delta x$ in \mbox{Eq. (\ref{AppendixEq})}. From that same equation, it is straightforward to observe that when the DE density vanishes, both $P_{\textrm{tot}}(x_{\dagger,ss})$ and $p_{\textrm{d}}(x_{\dagger,ss})$ remain finite and non-zero; therefore a $w$-singularity occurs at the transition point $x_{\dagger,ss}$. In fact, we find that the right and left limits of the EoS parameter at the moment of transition are give by $w_\mathrm{d,ss}^\pm\rightarrow\pm\infty$. As the EoS
parameter cannot be defined even for finite $\Delta x$. Nevertheless, a $w$-singularity is far milder than a sudden singularity \cite{Bouhmadi-Lopez:2019zvz}. All the results presented in this subsection, remains valid in absence of radiation.

\subsection{ECDM singularities}

Using Eq. (\ref{eosecdm}) we can compute the total pressure and the total density
\begin{equation}
\begin{aligned}
    \rho_{\textrm{tot}}= &\rho_{\textrm{r},0} e^{-4x}+ \rho_{\textrm{m},0} e^{-3x}+ \rho_{\textrm{d},0} \text{ Erf}[\eta(x-x_\dagger)], \\
     P_{\textrm{tot}}= & \frac{1}{3} \rho_{\textrm{r},0} e^{-4x}
     -\rho_{\textrm{d},0} \text{Erf}[\eta(x-x_\dagger)] \\
     & -\rho_{\textrm{d},0}\frac{2\eta}{3\sqrt{\pi}}e^{-\eta^2(x-x_\dagger)^2}.
\end{aligned}
\label{ecdmtot}
\end{equation}

By evaluating $\rho_{\textrm{tot}}(x_{\dagger,e})$ and $P_{\textrm{tot}}(x_{\dagger,e})$, we find
\begin{equation}
\begin{aligned}
    & \rho_{\textrm{tot}}(x_{\dagger,e})= \rho_{\textrm{r},0} e^{-4x_{\dagger,e}}+ \rho_{\textrm{m},0} e^{-3x_{\dagger,e}}, \\
    & P_{\textrm{tot}}(x_{\dagger,e})=\frac{1}{3}\rho_{\textrm{r},0} e^{-4x_{\dagger,e}}-\rho_{\textrm{d},0}\frac{2\eta}{3\sqrt{\pi}}.
\end{aligned}
\end{equation}

Notice that $\rho_{\textrm{tot}}(x_{\dagger,e})$ does not depend on $\eta$ and remains finite, whereas $P_{\textrm{tot}}(x_{\dagger,e})$ is negative and also finite for finite values of $\eta$. Nevertheless, in the limit $\eta \rightarrow \infty$, the DE density reduces to  $\Lambda_\textrm{s}$CDM. For such a limit
 \begin{equation}
     \lim_{\Delta x\rightarrow\infty}P_{\textrm{tot}}(x_{\dagger,e})=-\infty.
 \end{equation}

Thus a type II (sudden) cosmological singularity happens when the former limit is taken. On the contrary, for a finite $\Delta x$, it can be shown that a $w$-singularity occurs at the transition point.

\section{Cosmography in DE models}\label{appendixb}
In the current appendix, we present the general expression of the cosmographic parameters and the state finder hierarchy. To achieve this goal, we begin by rewriting the Friedmann equation as follows: 
\begin{equation}
    \dot{a}=e^x\left(\Omega_{\mathrm{r},0}e^{-4x}+\Omega_{\mathrm{m},0}e^{-3x}+\Omega_{\mathrm{d},0} \frac{\rho_\mathrm{d}(x)}{\rho_{\mathrm{d},0}}\right)^{1/2}
\end{equation}
where $x=\ln(a/a_0)$. It is helpful to write the cosmographic parameters in terms of $x$:
\begin{equation}
    \begin{aligned}
    A_2 &=\frac{\ddot{a}}{aH^2}=\frac{1}{aH}\frac{d\dot{a}}{dx},\\
    A_3 &=\frac{{a}^{(iii)}}{aH^3}=\frac{1}{aH^2}\frac{d}{dx}\left[H\frac{d\dot{a}}{dx}\right],\\
    A_4 &=\frac{{a}^{(iv)}}{aH^4}=\frac{1}{aH^3}\frac{d}{dx}\left[H\frac{d}{dx}\left[H\frac{d\dot{a}}{dx}\right]\right],\\
    A_5 &=\frac{{a}^{(v)}}{aH^5}=\frac{1}{aH^4}\frac{d}{dx}\left[H\frac{d}{dx}\left[H\frac{d}{dx}\left[H\frac{d\dot{a}}{dx}\right]\right]\right].
    \end{aligned}
\end{equation}

\vspace*{2ex}

Likewise, the statefinder hierarchy parameters read:

\begin{equation}
\begin{aligned}
    S_3^{(1)}(x) &=\frac{{a}^{(iii)}}{aH^3}=\frac{1}{aH^2}\frac{d}{dx}\left[H\frac{d\dot{a}}{dx}\right],\\
    S_4^{(1)}(x) &=\frac{{a}^{(iv)}}{aH^4}=\frac{1}{aH^3}\frac{d}{dx}\left[H\frac{d}{dx}\left[H\frac{d\dot{a}}{dx}\right]\right]\\
    & +3\left(1-\frac{1}{aH}\frac{d\dot{a}}{dx}\right),\\
    S_5^{(1)}(x) &=\frac{{a}^{(v)}}{aH^5}=\frac{1}{aH^4}\frac{d}{dx}\left[H\frac{d}{dx}\left[H\frac{d}{dx}\left[H\frac{d\dot{a}}{dx}\right]\right]\right]\\
    & -2\left(4-3\frac{1}{aH}\frac{d\dot{a}}{dx}\right)\left(1-\frac{1}{aH}\frac{d\dot{a}}{dx}\right),\\
    \tilde{s} & = \frac{1-\frac{1}{aH^2}\frac{d}{dx}\left[H\frac{d\dot{a}}{dx}\right]}{3\left(\frac{1}{aH}\frac{d\dot{a}}{dx}+\frac{1}{2}\right).}
\end{aligned}
\end{equation}

\begin{widetext}
\section{Cosmographic parameters for the analysed models}\label{appendixc}

The analytic solution for the cosmographic parameters of SSCDM will not be written as they are too lengthy and do not give  match input.
All the results in this section are presented for a spatially flat FLRW universe filled with DE, matter, and radiation.

\subsection{\texorpdfstring{$\Lambda\textrm{C}$}CDM}

\begin{align}
    q(x) &= - 1 + \frac{3  \Omega_{\mathrm{m},0} e^x + 4 \Omega_{\mathrm{r},0}}{2 \left(  \Omega_{\mathrm{d},0} e^{4x}  + \Omega_{\mathrm{m},0} e^x  + \Omega_{\mathrm{r},0} \right)} \label{eq:q}, \\
    j(x) &= 1 + \frac{2 \Omega_{\mathrm{r},0}}{ \Omega_{\mathrm{d},0} e^{4x} + \Omega_{\mathrm{m},0} e^x  + \Omega_{\mathrm{r},0}} \label{eq:j}, \\
    s(x) &= 1 - \frac{2 \Omega_{\mathrm{r},0} (3  \Omega_{\mathrm{m},0} e^x + 4 \Omega_{\mathrm{r},0}) + 
    3 ( \Omega_{\mathrm{d},0} e^{4x} +\Omega_{\mathrm{m},0}  e^x  + \Omega_{\mathrm{r},0}) (3  \Omega_{\mathrm{m},0} e^x + 8 \Omega_{\mathrm{r},0})}
    {2 ( \Omega_{\mathrm{d},0} e^{4x} + \Omega_{\mathrm{m},0}  e^x + \Omega_{\mathrm{r},0})^2} \label{eq:s}, \\
    l(x) &= \frac{2  \Omega_{\mathrm{d},0}^2 e^{8x}+ 10  \Omega_{\mathrm{d},0}  \Omega_{\mathrm{m},0}e^{5x} + 
    35  \Omega_{\mathrm{m},0}^2 e^{2x} + 60  \Omega_{\mathrm{d},0}  \Omega_{\mathrm{r},0} e^{4x}+ 
    210  \Omega_{\mathrm{m},0} \Omega_{\mathrm{r},0} e^x + 210 \Omega_{\mathrm{r},0}^2}
    {2 (\Omega_{\mathrm{d},0} e^{4x}  +\Omega_{\mathrm{m},0} e^x  + \Omega_{\mathrm{r},0})^2} \label{eq:l}.
\end{align}

\subsection{\texorpdfstring{$\Lambda_s\textrm{C}$}CDM}

\begin{align}
        q(x) &= - 1 + \frac{3  \Omega_{\mathrm{m},0} e^x + 4 \Omega_{\mathrm{r},0}}{2 \left(  \Omega_{\mathrm{d},0} \textrm{sgn}(z-z_{\dagger,s})e^{4x}  + \Omega_{\mathrm{m},0} e^x  + \Omega_{\mathrm{r},0} \right)}, \\
        j(x) &= 1 + \frac{2 \Omega_{\mathrm{r},0}}{ \Omega_{\mathrm{d},0} \textrm{sgn}(z-z_{\dagger,s}) e^{4x} + \Omega_{\mathrm{m},0} e^x  + \Omega_{\mathrm{r},0}}, \\
        s(x) &= 1 - \frac{2 \Omega_{\mathrm{r},0} (3  \Omega_{\mathrm{m},0} e^x + 4 \Omega_{\mathrm{r},0}) + 
    3 ( \Omega_{\mathrm{d},0}\textrm{sgn}(z-z_{\dagger,s}) e^{4x} +\Omega_{\mathrm{m},0}  e^x  + \Omega_{\mathrm{r},0}) (3  \Omega_{\mathrm{m},0} e^x + 8 \Omega_{\mathrm{r},0})}
    {2 ( \Omega_{\mathrm{d},0} \textrm{sgn}(z-z_{\dagger,s})e^{4x} + \Omega_{\mathrm{m},0}  e^x + \Omega_{\mathrm{r},0})^2}, \\
    l(x) &= \frac{2  \Omega_{\mathrm{d},0}^2 e^{8x}+ 10  \Omega_{\mathrm{d},0}  \Omega_{\mathrm{m},0} \textrm{sgn}(z-z_{\dagger,s})e^{5x} + 
    35  \Omega_{\mathrm{m},0}^2 e^{2x} + 60  \Omega_{\mathrm{d},0} \Omega_{\mathrm{r},0}\textrm{sgn}(z-z_{\dagger,s}) e^{4x} + 
    210  \Omega_{\mathrm{m},0} \Omega_{\mathrm{r},0} e^x + 210 \Omega_{\mathrm{r},0}^2}
    {2 (\Omega_{\mathrm{d},0}\textrm{sgn}(z-z_{\dagger,s}) e^{4x}  +\Omega_{\mathrm{m},0} e^x  + \Omega_{\mathrm{r},0})^2}.
\end{align}

\subsection{Ladder models (Le\texorpdfstring{$\Lambda\textrm{C}$}CDM and Lo\texorpdfstring{$\Lambda\textrm{C}$}CDM)}

We will refer to both the even, $\Lambda_{\mathrm{le}}(z)$, and odd, $\Lambda_{\mathrm{lo}}(z) $, cases as $\Lambda_{\mathrm{L}}(e^{-x}-1) $  for simplicity
\begin{align}
        q(x) &= - 1 + \frac{3  \Omega_{\mathrm{m},0} e^x + 4 \Omega_{\mathrm{r},0}}{2 \left(  \Omega_{\mathrm{d},0} \Lambda_\mathrm{L}(e^{-x}-1) e^{4x}  + \Omega_{\mathrm{m},0} e^x  + \Omega_{\mathrm{r},0} \right)}, \\
        j(x) &= 1 + \frac{2 \Omega_{\mathrm{r},0}}{ \Omega_{\mathrm{d},0} \Lambda_\mathrm{L}(e^{-x}-1) e^{4x} + \Omega_{\mathrm{m},0} e^x  + \Omega_{\mathrm{r},0}}, \\
        s(x) &= 1 - \frac{2 \Omega_{\mathrm{r},0} (3  \Omega_{\mathrm{m},0} e^x + 4 \Omega_{\mathrm{r},0}) + 
        3 ( \Omega_{\mathrm{d},0} \Lambda_\mathrm{L}(e^{-x}-1) e^{4x} +\Omega_{\mathrm{m},0}  e^x  + \Omega_{\mathrm{r},0}) (3  \Omega_{\mathrm{m},0} e^x + 8 \Omega_{\mathrm{r},0})}
        {2 ( \Omega_{\mathrm{d},0} \Lambda_\mathrm{L}(e^{-x}-1)e^{4x} + \Omega_{\mathrm{m},0}  e^x + \Omega_{\mathrm{r},0})^2}, \\
        l(x) &= \frac{2  \Omega_{\mathrm{d},0}^2   e^{8x}+ 10  \Omega_{\mathrm{d},0}   \Omega_{\mathrm{m},0} \Lambda_\mathrm{L}(e^{-x}-1) e^{5x}+ 
        35  \Omega_{\mathrm{m},0}^2 e^{2x} + 60  \Omega_{\mathrm{d},0} \Omega_{\mathrm{r},0} \Lambda_\mathrm{L}(e^{-x}-1) e^{4x}+ 
        210  \Omega_{\mathrm{m},0} \Omega_{\mathrm{r},0} e^x + 210 \Omega_{\mathrm{r},0}^2}
        {2 (\Omega_{\mathrm{d},0} \Lambda_\mathrm{L}(e^{-x}-1) e^{4x}  +\Omega_{\mathrm{m},0} e^x  + \Omega_{\mathrm{r},0})^2}.
\end{align}

\subsection{ECDM}
\begin{align}
    q(x) &= -1+\frac{-
    \Omega_{\mathrm{d},0} \frac{2}{\sqrt{\pi}} e^{4x - (x - x_{\dagger,e})^2 \eta^2} \eta
    +  3\Omega_{\mathrm{m},0} e^x + 4 \Omega_{\mathrm{r},0}}
    {2 \left( \Omega_{\mathrm{m},0}  e^x+ \Omega_{\mathrm{r},0} +  \Omega_{\mathrm{d},0} e^{4x} \operatorname{Erf}[\eta(x - x_{\dagger,e})] \right)}
    \label{eq:qe}, \\
    j(x) &= 1+
    \frac{e^{-(x - x_{\dagger,e})^2 \eta^2} \left[
    - \Omega_{\mathrm{d},0} e^{4x} \eta (-3 + 2(x - x_{\dagger,e}) \eta^2)  
    +2\sqrt{\pi} \Omega_{\mathrm{r},0}
    \right]}
    {\sqrt{\pi} \left( \Omega_{\mathrm{m},0} e^x + \Omega_{\mathrm{r},0} +  \Omega_{\mathrm{d},0} e^{4x}\operatorname{Erf}[\eta(x - x_{\dagger,e})] \right)}
    \label{eq:je}, \\
    s(x) &= \left[2\pi \left( \Omega_{\mathrm{m},0} e^{x} + \Omega_{\mathrm{r},0} + \Omega_{\mathrm{d},0} e^{4x} \operatorname{Erf}[\eta(x - x_{\dagger,e})] \right)^2\right]^{-1} \Biggl\{
    e^{-2 (x - x_{\dagger,e})^2 \eta^2} \left[
    -2 e^{8x} \eta^2 (-3 + 2(x - x_{\dagger,e}) \eta^2) \Omega_{\mathrm{d},0}^2 \right. \nonumber \\
    & \left.
    + e^{4x + (x - x_{\dagger,e})^2 \eta^2} \sqrt{\pi} \eta \Omega_{\mathrm{d},0}
    \left( e^x (3 - 2 (2 + 5x - 5x_{\dagger,e}) \eta^2 + 8 (x - x_{\dagger,e})^2 \eta^4) \Omega_{\mathrm{m},0}
    + 4 (1 + (-1 - 2x + 2x_{\dagger,e}) \eta^2 \right.\right. \nonumber \\
    & \left.\left.
    + 2 (x - x_{\dagger,e})^2 \eta^4) \Omega_{\mathrm{r},0} \right)
    - e^{2(x - x_{\dagger,e})^2 \eta^2} \pi \left(7 e^{2x} \Omega_{\mathrm{m},0}^2 + 35 \Omega_{\mathrm{m},0} e^{x} \Omega_{\mathrm{r},0} + 30 \Omega_{\mathrm{r},0}^2 \right) \right. \nonumber \\
    & \left.+ e^{4x + (x - x_{\dagger,e})^2 \eta^2} \sqrt{\pi} \Omega_{\mathrm{d},0} \operatorname{Erf}[\eta(x - x_{\dagger,e})]
    \left(4 e^{4x} \eta (3 + (-1 - 4x + 4x_{\dagger,e}) \eta^2 
    + 2 (x - x_{\dagger,e})^2 \eta^4) \Omega_{\mathrm{d},0}
    \right.\right. \nonumber \\
    & \left.\left. + e^{(x - x_{\dagger,e})^2 \eta^2} \sqrt{\pi} (-5 (\Omega_{\mathrm{m},0} e^{x} + 4 \Omega_{\mathrm{r},0}) + 2 \Omega_{\mathrm{d},0} e^{4x} \operatorname{Erf}[\eta(x - x_{\dagger,e})]) \right)
    \right]\Biggl\}
    \label{eq:ss}, \\
    l(x) &= \left[2\pi \left( \Omega_{\mathrm{m},0} e^{x} + \Omega_{\mathrm{r},0} + \Omega_{\mathrm{d},0} e^{4x} \operatorname{Erf}[\eta(x - x_{\dagger,e})] \right)^2\right]^{-1}\Biggl\{
    e^{-2 (x - x_{\dagger,e})^2 \eta^2} \left[
    2 e^{8x} \eta^2 (15 - 6(1 + 5x - 5x_{\dagger,e}) \eta^2 \right. \nonumber \\
    & \left. + 16 (x - x_{\dagger,e})^2 \eta^4) \Omega_{\mathrm{d},0}^2
    - 2 e^{4x + (x - x_{\dagger,e})^2 \eta^2} \sqrt{\pi} \eta \Omega_{\mathrm{d},0}
    \left( e^x (8 + (1 - 7x + 7x_{\dagger ,e}) \eta^2 - 2 (x - x_{\dagger ,e}) (6 + x - x_{\dagger,e}) \eta^4 \right.\right. \nonumber \\
    & \left.\left.
    + 8 (x - x_{\dagger ,e})^3 \eta^6) \Omega_{\mathrm{m},0} + 2 (13 + (-1 - 4x + 4x_{\dagger ,e}) \eta^2
    + 2 (-3 + x - x_{\dagger ,e}) (x - x_{\dagger ,e}) \eta^4 + 4 (x - x_{\dagger ,e})^3 \eta^6) \Omega_{\mathrm{r},0} \right) \right. \nonumber \\
    & \left.
    + 35 e^{2(x - x_{\dagger ,e})^2 \eta^2} \pi (e^{2x} \Omega_{\mathrm{m},0}^2 + 6 \Omega_{\mathrm{m},0} e^{x} \Omega_{\mathrm{r},0} + 6 \Omega_{\mathrm{r},0}^2)
    + 2 e^{4x + (x - x_{\dagger ,e})^2 \eta^2} \sqrt{\pi} \Omega_{\mathrm{d},0} \operatorname{Erf}[(x - x_{\dagger ,e}) \eta]\times \right. \nonumber \\
    & \left.
    \left(-2 e^{4x} \eta (-5 + 5(1 + 2x - 2x_{\dagger ,e}) \eta^2 - 2 (3 + 5x - 5x_{\dagger ,e}) (x - x_{\dagger ,e}) \eta^4
    + 4 (x - x_{\dagger ,e})^3 \eta^6) \Omega_{\mathrm{d},0} \right.\right. \nonumber \\
    & \left.\left.+ e^{(x - x_{\dagger ,e})^2 \eta^2} \sqrt{\pi} (5 \Omega_{\mathrm{m},0} e^{x}
    + 30 \Omega_{\mathrm{r},0} + \Omega_{\mathrm{d},0} e^{4x} \operatorname{Erf}[\eta(x - x_{\dagger ,e}) ]) \right)
    \right]\Biggl\}.
    \label{eq:ll}
\end{align}

\section{Statefinder parameters for the analysed models}
\label{appendixd}

The analytic solution for the statefinder parameters of SSCDM will not be written as they are too lengthy and do not give  match input.
All the results in this section are presented for a spatially flat FLRW universe filled with DE, matter, and radiation.

\subsection{\texorpdfstring{$\Lambda\textrm{C}$} CDM}

\begin{align}
S_3^{(1)}& =  1+\frac{2 \Omega_{\mathrm{r},0} e^{-x}}{\Omega_{\mathrm{r}, 0} e^{-x}+\Omega_{\mathrm{m}, 0}+\Omega_{\mathrm{d}, 0}e^{-3x}},\\
S_4^{(1)} & =1-\left[\Omega_{\mathrm{r} 0} e^{-x}+\Omega_{\mathrm{m}, 0}+\Omega_{\mathrm{d}, 0}e^{-3x}\right]^{-2}\left[\left( 10 \Omega_{\mathrm{r}, 0} e^{-x}+9 \Omega_{\mathrm{m}, 0}+6 \Omega_{\mathrm{d}, 0}e^{-3x}\right)\Omega_{\mathrm{r}, 0}e^{-x}\right] , \\
S_5^{(1)} & =1+\left[\Omega_{\mathrm{r}, 0} e^{-x}+\Omega_{\mathrm{m}, 0}+\Omega_{\mathrm{d}, 0}e^{-3x}\right]^{-2}\left\{\left[ 76 \Omega_{\mathrm{r}, 0} e^{-x}+60 \Omega_{\mathrm{m}, 0} + 57 \Omega_{\mathrm{d}, 0}e^{-3x}\right] \Omega_{\mathrm{r}, 0} e^{-x} \right\} , \\
\tilde{s} & =  \frac{4 \Omega_{\mathrm{r}, 0}e^{-x}}{3 \Omega_{\mathrm{r}, 0} e^{-x}-9  \Omega_{\mathrm{d}, 0}e^{-3x}} .
\end{align}

\subsection{\texorpdfstring{$\Lambda_s\textrm{C}$}CDM}

\begin{align}
S_3^{(1)} =  &1+\frac{2 \Omega_{\mathrm{r},0} e^{-x}}{\Omega_{\mathrm{r}, 0} e^{-x}+\Omega_{\mathrm{m}, 0}+\Omega_{\mathrm{d}, 0}\textrm{sgn}(z-z_{\dagger ,s})e^{-3x}},\\
S_4^{(1)}  = &1-\left[\Omega_{\mathrm{r} 0} e^{-x}+\Omega_{\mathrm{m}, 0}+\Omega_{\mathrm{d}, 0}\textrm{sgn}(z-z_{\dagger ,s})e^{-3x}\right]^{-2}\left[\left( 10 \Omega_{\mathrm{r}, 0} e^{-x} \right.\right. \nonumber \\
& \left.\left.+9 \Omega_{\mathrm{m}, 0}+6 \Omega_{\mathrm{d}, 0}\textrm{sgn}(z-z_{\dagger ,s})e^{-3x}\right)\Omega_{\mathrm{r}, 0}e^{-x}\right] , \\
S_5^{(1)}  = &1+\left[\Omega_{\mathrm{r}, 0} e^{-x}+\Omega_{\mathrm{m}, 0}+\Omega_{\mathrm{d}, 0}\textrm{sgn}(z-z_{\dagger ,s})e^{-3x}\right]^{-2}\left\{\left[ 76 \Omega_{\mathrm{r}, 0} e^{-x} \right.\right. \nonumber \\
& \left.\left. +60 \Omega_{\mathrm{m}, 0} + 57 \Omega_{\mathrm{d}, 0}\textrm{sgn}(z-z_{\dagger ,s})e^{-3x}\right] \Omega_{\mathrm{r}, 0} e^{-x} \right\} , \\
\tilde{s} & =  \frac{4 \Omega_{\mathrm{r}, 0}e^{-x}}{3 \Omega_{\mathrm{r}, 0} e^{-x}-9  \Omega_{\mathrm{d}, 0}\textrm{sgn}(z-z_{\dagger ,s})e^{-3x}} .
\end{align}

\subsection{Ladder models (Le\texorpdfstring{$\Lambda\textrm{C}$}CDM and Lo\texorpdfstring{$\Lambda\textrm{C}$}CDM)}
We will refer to both the even, $\Lambda_{\mathrm{le}}(z)$, and odd, $\Lambda_{\mathrm{lo}}(z) $, cases as $\Lambda_{\mathrm{L}}(e^{-x}-1) $  for simplicity
\begin{align}
S_3^{(1)}& =  1+\frac{2 \Omega_{\mathrm{r},0} e^{-x}}{\Omega_{\mathrm{r}, 0} e^{-x}+\Omega_{\mathrm{m}, 0}+\Omega_{\mathrm{d}, 0} \Lambda_\mathrm{L}(e^{-x}-1)e^{-3x}},\\
S_4^{(1)} & =1-\left[\Omega_{\mathrm{r} 0} e^{-x}+\Omega_{\mathrm{m}, 0}+\Omega_{\mathrm{d}, 0} \Lambda_\mathrm{L}(e^{-x}-1)e^{-3x}\right]^{-2}\left[\left( 10 \Omega_{\mathrm{r}, 0} e^{-x}+9 \Omega_{\mathrm{m}, 0}+6 \Omega_{\mathrm{d}, 0} \Lambda_\mathrm{L}(e^{-x}-1)e^{-3x}\right)\Omega_{\mathrm{r}, 0}e^{-x}\right] , \\
S_5^{(1)} & =1+\left[\Omega_{\mathrm{r}, 0} e^{-x}+\Omega_{\mathrm{m}, 0}+\Omega_{\mathrm{d}, 0} \Lambda_\mathrm{L}(e^{-x}-1)e^{-3x}\right]^{-2}\left\{\left[ 76 \Omega_{\mathrm{r}, 0} e^{-x}+60 \Omega_{\mathrm{m}, 0} + 57 \Omega_{\mathrm{d}, 0} \Lambda_\mathrm{L}(e^{-x}-1)e^{-3x}\right] \Omega_{\mathrm{r}, 0} e^{-x} \right\} , \\
\tilde{s} & =  \frac{4 \Omega_{\mathrm{r}, 0}e^{-x}}{3 \Omega_{\mathrm{r}, 0} e^{-x}-9  \Omega_{\mathrm{d}, 0} \Lambda_\mathrm{L}(e^{-x}-1)e^{-3x}} .
\end{align}

\subsection{ECDM}

\begin{align}
S_3^{(1)} &= 1+\frac{e^{-(x - x_{\dagger,e})^2 \eta^2} (-e^{4x} \eta (-3 + 2 (x - x_{\dagger,e}) \eta^2) \Omega_{\mathrm{d},0}) +  2\sqrt{\pi}   \Omega_{\mathrm{r},0} }{\sqrt{\pi} (\Omega_{\mathrm{m},0} e^{x} + \Omega_{\mathrm{r},0} + \Omega_{\mathrm{d},0} e^{4x} \operatorname{Erf}[(x - x_{\dagger,e}) \eta])}, 
\\
S_4^{(1)} &= \frac{1}{2\pi (\Omega_{\mathrm{m},0} e^{x} + \Omega_{\mathrm{r},0} + \Omega_{\mathrm{d},0} e^{4x} \operatorname{Erf}[(x - x_{\dagger,e}) \eta])^2} e^{-2 (x - x_{\dagger,e})^2 \eta^2} \bigg(-2 e^{8x} \eta^2 (-3 + 2 (x - x_{\dagger,e}) \eta^2) \Omega_{\mathrm{d},0}^2 \nonumber \\
&\quad + e^{4x + (x - x_{\dagger,e})^2 \eta^2} \sqrt{\pi} \eta \Omega_{\mathrm{d},0} (e^x (-3 - 2 (2 + 5x - 5x_{\dagger,e}) \eta^2 + 8 (x - x_{\dagger,e})^2 \eta^4) \Omega_{\mathrm{m},0} \nonumber\\
&\quad + 2 (-1 + (-2 - 4x + 4x_{\dagger,e}) \eta^2 + 4 (x - x_{\dagger,e})^2 \eta^4) \Omega_{\mathrm{r},0}) \nonumber\\
&\quad + 2 e^{2 (x - x_{\dagger,e})^2 \eta^2} \pi (e^{2x} \Omega_{\mathrm{m},0}^2 - 7 \Omega_{\mathrm{m},0} e^{x} \Omega_{\mathrm{r},0} - 9 \Omega_{\mathrm{r},0}^2) \nonumber\\
&\quad + 2 e^{4x + (x - x_{\dagger,e})^2 \eta^2} \sqrt{\pi} \Omega_{\mathrm{d},0} \operatorname{Erf}[(x - x_{\dagger,e}) \eta] (e^{4x} \eta (3 + (-2 - 8x + 8x_{\dagger,e}) \eta^2 + 4 (x - x_{\dagger,e})^2 \eta^4) \Omega_{\mathrm{d},0} \nonumber\\
&\quad + e^{(x - x_{\dagger,e})^2 \eta^2} \sqrt{\pi} (2 \Omega_{\mathrm{m},0} e^{x} - 4 \Omega_{\mathrm{r},0} + \Omega_{\mathrm{d},0} e^{4x} \operatorname{Erf}[(x - x_{\dagger,e}) \eta]))),
\\
S_5^{(1)} &= \frac{1}{\pi (\Omega_{\mathrm{m},0} e^{x} + \Omega_{\mathrm{r},0} + \Omega_{\mathrm{d},0} e^{4x} \operatorname{Erf}[(x - x_{\dagger,e}) \eta])^2} e^{-2 (x - x_{\dagger,e})^2 \eta^2} \bigg(e^{8x} \eta^2 (9 - 6 (1 + 5x - 5x_{\dagger,e}) \eta^2 \nonumber\\
&\quad + 16 (x - x_{\dagger,e})^2 \eta^4) \Omega_{\mathrm{d},0}^2 + e^{4x + (x - x_{\dagger,e})^2 \eta^2} \sqrt{\pi} \eta \Omega_{\mathrm{d},0} (-e^x (-12 + (1 - 7x + 7x_{\dagger,e}) \eta^2 \nonumber\\
&\quad - 2 (x - x_{\dagger,e}) (6 + x - x_{\dagger,e}) \eta^4 + 8 (x - x_{\dagger,e})^3 \eta^6) \Omega_{\mathrm{m},0} + 2 \eta^2 (1 + 4x - 4x_{\dagger,e} - 2 (-3 + x - x_{\dagger,e}) (x - x_{\dagger,e}) \eta^2 \nonumber\\
&\quad - 4 (x - x_{\dagger,e})^3 \eta^4) \Omega_{\mathrm{r},0}) + e^{2 (x - x_{\dagger,e})^2 \eta^2} \pi (e^{2x} \Omega_{\mathrm{m},0}^2 + 62 \Omega_{\mathrm{m},0} e^{x} \Omega_{\mathrm{r},0} + 77 \Omega_{\mathrm{r},0}^2) \nonumber\\
&\quad + e^{4x + (x - x_{\dagger,e})^2 \eta^2} \sqrt{\pi} \Omega_{\mathrm{d},0} \operatorname{Erf}[(x - x_{\dagger,e}) \eta] (-2 e^{4x} \eta (-6 + 5 (1 + 2x - 2x_{\dagger,e}) \eta^2 \nonumber\\
&\quad - 2 (3 + 5x - 5x_{\dagger,e}) (x - x_{\dagger,e}) \eta^4 + 4 (x - x_{\dagger,e})^3 \eta^6) \Omega_{\mathrm{d},0} + e^{(x - x_{\dagger,e})^2 \eta^2} \sqrt{\pi} (2 \Omega_{\mathrm{m},0} e^{x} + 26 \Omega_{\mathrm{r},0} \nonumber\\
&\quad + \Omega_{\mathrm{d},0} e^{4x} \operatorname{Erf}[(x - x_{\dagger,e}) \eta]))) ,
\\
\tilde{s} &= \frac{2 e^{4x} \eta (-3 + 2 (x - x_{\dagger,e}) \eta^2) \Omega_{\mathrm{d},0} - 4 e^{(x - x_{\dagger,e})^2 \eta^2} \sqrt{\pi} \Omega_{\mathrm{r},0}}{6 e^{4x} \eta \Omega_{\mathrm{d},0} - 3 e^{(x - x_{\dagger,e})^2 \eta^2} \sqrt{\pi} (\Omega_{\mathrm{r},0} - 3 \Omega_{\mathrm{d},0} e^{4x} \operatorname{Erf}[(x - x_{\dagger,e}) \eta])}.
\end{align}

\end{widetext}

\bibliography{bibliography}

\end{document}